\documentclass{aa}  

\usepackage{graphicx}
\usepackage{natbib}
\usepackage{txfonts}
\usepackage{textcomp}
\pdfoutput=1
\bibpunct{(}{)}{;}{a}{}{,}             
\catcode`\@=11
\def\gsim{\ifmmode{\mathrel{\mathpalette\@versim>}}
    \else{$\mathrel{\mathpalette\@versim>}$}\fi}
\def\lsim{\ifmmode{\mathrel{\mathpalette\@versim<}}
    \else{$\mathrel{\mathpalette\@versim<}$}\fi}
\def\@versim#1#2{\lower 2.9truept \vbox{\baselineskip 0pt \lineskip
    0.5truept \ialign{$\m@th#1\hfil##\hfil$\crcr#2\crcr\sim\crcr}}}
\catcode`\@=12

\begin{document} 

 \title{The GIRAFFE Inner Bulge Survey (GIBS) III. Metallicity
distributions and kinematics of 26 Galactic bulge fields
   \thanks{Based on observations taken with ESO telescopes at 
   the La Silla Paranal Observatory under programme IDs 71.B-0617,
   385.B-0735 and 187.B-0909}}

\author{
M. Zoccali\inst{1,2} 
 \and
S. Vasquez\inst{1,2}
 \and
O. A. Gonzalez\inst{3}
 \and           
E. Valenti\inst{4}
\and
A. Rojas-Arriagada\inst{2,1}
 \and
J. Minniti\inst{1,2}
 \and
M. Rejkuba\inst{4,5}
 \and
D. Minniti\inst{6,2,7}
 \and 
A. McWilliam\inst{8}   
 \and 
C. Babusiaux\inst{9}   
 \and
V. Hill\inst{10}
\and 
A. Renzini\inst{11}
}

\institute{
Instituto de Astrof\'isica, Pontificia Universidad  Cat\'olica de Chile, 
Av. Vicu\~na Mackenna 4860, 782-0436 Macul, Santiago, Chile\\
\email{mzoccali@astro.puc.cl}
\and
Millennium Institute of Astrophysics, Av. Vicu\~na Mackenna 4860, 
782-0436 Macul, Santiago, Chile
\and
UK Astronomy Technology Centre, Royal Observatory, Blackford Hill, 
Edinburgh, UK
\and
European Southern Observatory, Karl-Schwarzschild Strasse 2, D-85748 
Garching, Germany
\and
Excellence Cluster Origin and Structure of the Universe, Boltzmannstr. 2, 
D-85748 Garching bei München, Germany
Departamento de Ciencias Físicas, Universidad Andrés Bello, 220 Rep\'ublica, 
Santiago, Chile
\and
Departamento de Fisica, Facultad de Ciencias Exactas, Universidad Andres Bello
Av. Fernandez Concha 700, Las Condes, Santiago, Chile
\and
Vatican Observatory, V00120 Vatican City State, Italy
\and
The Observatories of the Carnegie Institution of Washington, 
813 Santa Barbara St., Pasadena, CA 91101--1292
\and
GEPI, Observatoire de Paris, CNRS UMR 8111, Universit\'e Paris Diderot, 
F-92125, Meudon, Cedex, France
\and
Universit\'e de la C\^ote d'Azur, Observatoire de la C\^ote d'Azur, CNRS, 
Laboratoire Lagrange, Bd de l'Observatoire, CS 34229, 06304 Nice Cedex 4, France
\and
INAF - Osservatorio Astronomico di Padova, vicolo dell'Osservatorio 5, 35122, 
Padova, Italy
}


  \abstract
{Several recent  studies have demonstrated  that the Galactic bulge  hosts two
  components with different mean metallicities, and possibly different spatial
  distribution  and  kinematics.   As  a  consequence,  both  the  metallicity
  distribution and  the radial velocity  of bulge stars vary  across different
  line of sights.}
{We present here  the metallicity distribution function of red  clump stars in
  26 fields spread across  a wide area of the bulge,  with special emphasis on
  fields close to  Galactic plane, at latitudes  $b=-2^\circ$ and $b=-1^\circ$,
  that were not explored before.}
{This paper  includes new metallicities  from a  sample of $\sim$5000  K giant
  stars,  observed  at spectral  resolution  R$\sim$6500,  in the  Calcium  II
  Triplet  region. They  are  the  main dataset  of  the  GIRAFFE Inner  Bulge
  Survey. As part of the same  survey we have previously published results for
  a sample  of $\sim$600 K  giant stars, at latitude  $b\sim-4^\circ$, derived
  from higher resolution spectra (R=22,500).}
{The combined  sample allows us to  trace and characterize the  metal poor and
  metal rich bulge  populations down to the inner bulge.  We present a density
  map  for each  of the  two components.   Contrary to  the expectations  from
  previous works,  we found  the metal  poor population  to be  more centrally
  concentrated than the  metal rich one, and with a  more axisymmetric spatial
  distribution.  The metal rich population, on  the other hand, is arranged in
  a  boxy  distribution,   consistent  with  an  edge-on   bar.   By  coupling
  metallicities and radial  velocities we show that the  metal poor population
  has a velocity  dispersion that varies rather mildly with  latitude.  On the
  contrary, the metal  rich population has a low velocity  dispersion far from
  the plane  ($b=-8.5^\circ$), but  it has a  steeper gradient  with latitude,
  becoming  higher   than  the   metal  poor  one   in  the   innermost  field
  ($b=-1^\circ$).  }
{This work provides new observational constraints on the actual chemodynamical 
properties of the Galactic bulge, that will help discriminating among different
formation models.}


\titlerunning{GIBS-III: MDFs and kinematics for 26 fields}
\authorrunning{Zoccali et al.}

   \maketitle

\section{Introduction}

Until a decade ago, our knowledge about the properties of the Galactic bulge,
the second most  massive component of the Milky Way,  was relatively poor. Yet
we thought  we knew most  of what  was relevant. We  knew it was  an elongated
spheroid   \citep[i.e.,  a   bar][]{stanek+94},  with   a  broad   metallicity
distribution    \citep[][]{rich+88},     containing    mostly     old    stars
\citep[][]{ortolani+95}. The  results of  the investigations cited  above were
refined by  several other authors \citep[e.g.][]{mwr+94,  zoccali+03} but none
of them covered a wide area of the  bulge, thus the common belief was that the
bulge was a rather uniform population.

Wide area photometric and spectroscopic surveys conducted in the last 10 years
revealed that the bulge is much more complex. 

Concerning the  bulge three dimensional  structure, we  now know that  the bar
flares  up in  a  boxy/peanut, or  X-shape \citep[][]{mcwilliam+10,  nataf+10,
  saito+11, wegg+13},  unequivocal signature of a  formation scenario starting
from  a disk,  whose  dynamical  instabilities funnel  stars,  and maybe  gas,
towards the center forming a bar, which later bends and buckles giving rise to
the X-shape  \citep[][]{patsis+02,athanassoula+05}.  However we also  know that
the oldest  and more  metal poor stars  do not follow  the same  structure. In
fact,  RR Lyrae  trace a  more axisymmetric  component, compared  with the  RC
\citep[][]{dekany+13, pietrukowicz+15}.  Also, metal poor red clump (RC) stars
do   not   trace    the   X-shape   \citep[][]{ness+12,   rojas-arriagada+14}.
Furthermore,  there is  evidence for  an axysimmetric  structure within  the
inner 250pc  of the bulge  \citep[][]{gonzalez+11bar, gerhard+12} and for  a thin
extension  of  the bar,  called  the  {\it  long} bar  \cite[][and  references
  therein]{wegg+15}. A recent review of the 3D structure of the Galactic bulge
can be found in \cite[][]{zoccali+16}.

Another fundamental  characteristic of any  stellar system is  the metallicity
distribution function (MDF); indeed, stellar  systems may be differentiated by
their mean metallicity, or [Fe/H], and  by the metallicity dispersion.  At any
moment,  the  gas-phase  metal  content   consists  of  the  integral  of  the
nucleosynthetic yields of the preceeding  history of star formation, resulting
in an  increase of the mean  metallicity with time. The  observed MDF provides
constraints  on models  of  chemical  evolution.  For  example,  the low  mean
metallicity of stars  in the Galactic halo led  \citet{hartwick76} to conclude
that the halo  must have lost its gas, through  outflows, before enrichment to
higher metallicity  could occur.  On the  other hand, the  lack of  metal poor
stars  in  the solar  neighborhood,  relative  to  a  Closed Box  of  chemical
enrichment,  is   an  evidence  of   gas  inflow  during   chemical  evolution
\citep[e.g.,][]{pagel89}. Additionally, the mean MDF may  reveal radial or 
vertical gradients, possibly indicating  viscous  flows,  the  radial action  
of  bars,  or  dissipational collapse.  

As mentioned above,  the fact that the bulge  has a broad MDF was  known for a
long time. However only  in the last $\sim$15 years has  the measured MDF been
precise enough  to allow some  important conclusions on the  bulge properties.
\citet[][]{zoccali+08,  hill+11, ness+13mdf,  rojas-arriagada+14, gonzalez+15}
demonstrated  that  the bulge  MDF  is  bimodal,  with  the two  peaks  having
metallicity a  few dex below and  a few dex above  solar, respectively.  While
the mean  metallicity changes across the  bulge area, the position  of the two
peaks does not, thus the mean metallicity gradient, also known for a long time
from  photometry  or  low resolution  spectroscopy  \citep[e.g.][]{terndrup88,
  minniti+95}  is in  fact due  to a  different relative  fraction of  the two
metallicity components.  None of the studies mentioned above observed stars at
latitudes $|b|<4^\circ$.   Due to  the higher extinction  close to  the plane,
previous  studies of  metallicitites in  the inner  bulge ($|b|<4^\circ$)  are
either based  on very  low number  statistics, i.e., one  or two  dozens stars
\citep[e.g.,][]{rich+07, rich+12,  schultheis+15}, or on $\sim$100  stars with
poorer     metallicities,    compared     with     those    presented     here
\citep[][]{babusiaux+14}.

Interestingly,  the  bulge  metal-rich  and  metal-poor  sub-populations  have
vertical  scale  heights  reminiscent  of  the local  thin  and  thick  disks,
respectively, and  consistent with  the [$\alpha$/Fe] ratios  of the  thin and
thick disks \citep[][]{alves-brito+10, hill+11, gonzalez+11_alphas}.  However,
these bulge sub-populations have $\sim$0.4  dex higher [Fe/H] than the average
thin and  thick disk stars  in the solar  neighborhood.  Thus, if  these bulge
sub-populations  really represent  the inner  thin and  thick disks,  a radial
[Fe/H]   gradient   near  $-$0.05   dex/kpc   is   implied  for   both   disks
\citep[][]{mcwilliam+16}.  If the bulge was  built by the growth and buckling
of a  stellar bar,  then inner thin  and thick disk  stars entrained  into the
bulge  must  have retained  vertical  scale  heights characteristic  of  their
origin, and  resulting in the  present-day vertical metallicity  gradient. The
MDF and the chemical composition of bulge stars have been recently reviewed by
\citet[][]{ness+16, mcwilliam+16}.

Finally,  an observational  tool that  can be  critical to  discriminate among
different formation scenarios is the stellar kinematics. Indeed, galactic bars
usually present streaming  motions along the major axis, due  to the elongated
orbits that sustain the bar. For the  same reason, the phase space of stars in
the  bar shows  a significant  vertex deviation,  contrary to  the stars  in a
spheroidal component, showing  a more isotropic velocity  distribution, with a
higher   dispersion.   Although   \citet[][]{saha+13}   demonstrated  that   a
pre-existing spheroid would increase its rotation  velocity if a bar is formed
at  later  times,  the  other kinematical  differences  mentioned  above  help
discriminate  between two  or more  populations  if they  belong to  different
spatial structures.  The interested reader  may refer  to the review  on bulge
kinematics   by   \citet[][]{babusiaux+16}.    The   most   relevant,   recent
investigations  demonstrated   that  the  bulge  shows   cylindrical  rotation
\citep[][]{howard+09,  ness+13kin,  zoccali+14},  with a  velocity  dispersion
increasing towards the  Galactic plane, including a peak in  the inner $\sim$2
degrees from the Galactic center \citep[][hereafter Paper~I]{zoccali+14}.  The
kinematical properties of  the metal poor and metal rich  bulge components are
slightly different, with the metal  poor having higher velocity dispersion, at
least in the outer bulge \citep[][]{ness+13kin}.

We  present here  the MDF  derived for 26  bulge fields,  observed within  the
GIRAFFE Inner Bulge Survey (GIBS), an  ESO Large Programme (ID 187.B-0909; PI:
Zoccali) carried  out with the  GIRAFFE spectrograph of the  FLAMES instrument
\citep{pasquini+02} at the  Very Large Telescope (VLT). The aim  of the survey
is to investigate  how the metallicity and radial  velocity distributions vary
across  the bulge  area, with  special attention  to the  fields close  to the
Galactic  plane, at  $b=-2$  and $b=-1$,  not explored  by  other surveys.  We
combine  here the  metallicities of  the stars  with their  radial velocities,
already  discussed   in  Paper~I,  in   order  to  characterize   the  spatial
distribution and kinematics of the two bulge metallicity components.


\section{The data}

\begin{figure}
\centering
    {\includegraphics[angle=0,width=0.9\columnwidth]{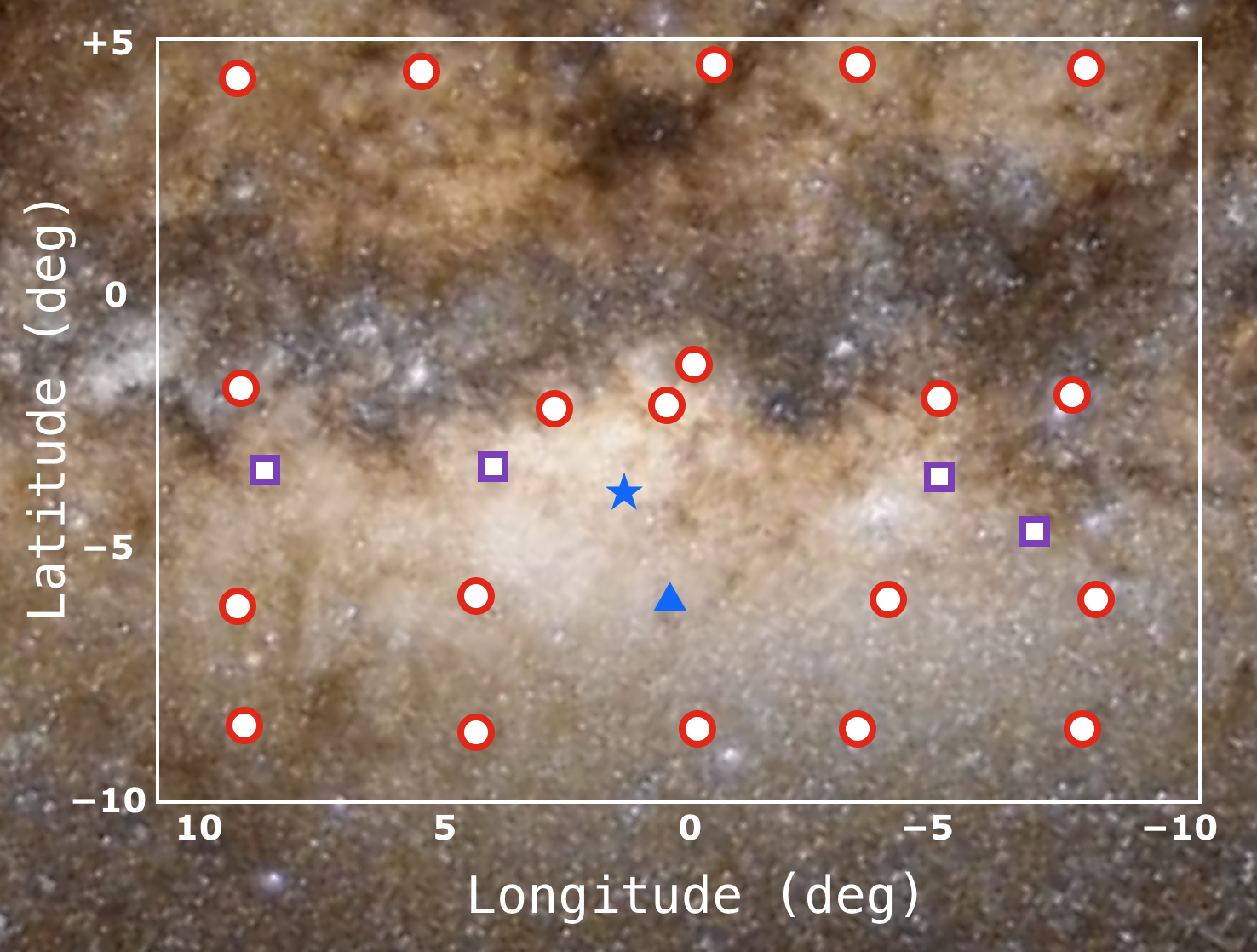}}
    \caption{Position in  the sky of  the 26  GIBS fields presented  here. The
      large  square is  the area  covered by  the VVV  Survey. Red  circle are
      fields  observed  at low  resolution  (LR8),  purple squares  are  field
      observed at  high resolution \citep[HR13][;  Paper~II]{gonzalez+15}. The
      blue  star is  the Baade's  Window field  used for  the CaT  calibration
      \citep[][]{vasquez+15},  and the  blue  triangle is  the ($0,-6$)  field
      discussed in \citet[][]{zoccali+08}.}
    \label{fields}
\end{figure}

The    present    investigation    is    based    on    spectra    for
$\sim$5500\footnote{The kinematical analysis  presented in Paper~I was
based on  $\sim$6500 stars, because  it included  archive spectra for  5 extra
fields at  $b=-2^\circ$ (from  Prog.  ID 089.B-0830)  and a  different, larger
sample at ($0,-6$), from \citet[][]{vasquez+15}. All these spectra had a lower
S/N, good  for radial velocity  measurement, but not for  reliable metallicity
measurements.}  RC stars in several fields across the Galactic bulge, as shown
in Fig.~\ref{fields}.  All of them are located within the region mapped by the
VVV Survey \citep[$-10<l<+10$ and $-10<b<+5$;][]{minniti+10}.

\begin{figure}[h]
\centering
    {\includegraphics[angle=0,height=8cm,width=8cm]{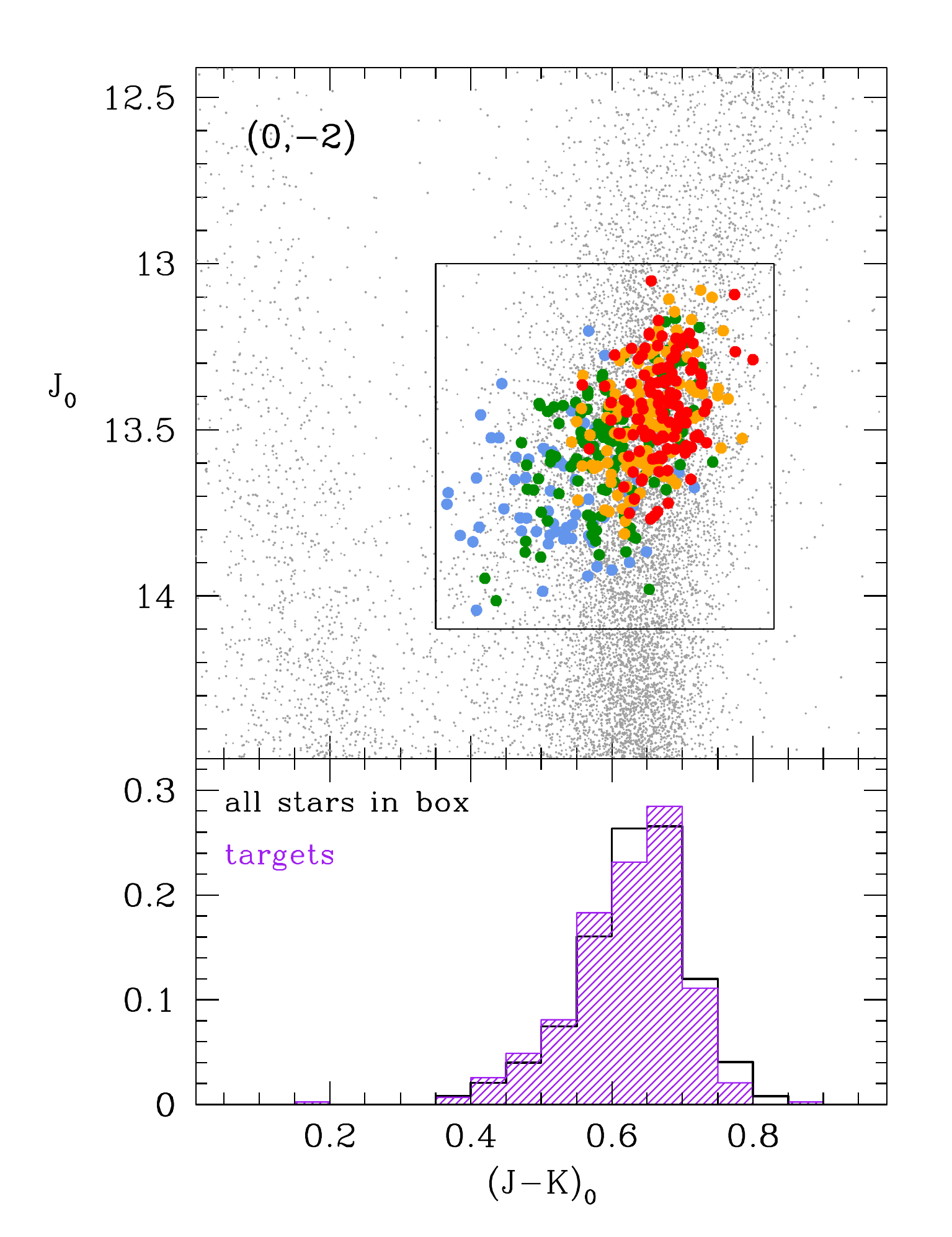}}
    \caption{Example  of the  GIBS target  selection criteria  for the  LRp0m1
      field.  Top:  the VVV CMD together  with the 432 GIRAFFE  targets, color
      coded according  to their  metallicity (blue: [Fe/H]$<-0.5$  dex; green:
      $-0.5<$[Fe/H]$<-0.1$;    orange:     $-0.1<$[Fe/H]$<+0.3$    and    red:
      [Fe/H]$>+0.3$.). Bottom: color histogram of  all the stars in the FLAMES
      field, in the black box of the CMD, compared with the color histogram of
      the selected targets.}
    \label{targets}
\end{figure}

The targets were selected from the  near infrared CMDs obtained within the VVV
Survey,   and   calibrated   to   the   2MASS   photometric   system   as   in
\citet[][]{gonzalez+11redd}.  The  target selection, discussed in  Paper~I, is
shown in Fig.~\ref{targets}: the GIRAFFE targets are RC stars, in a relatively
narrow range of apparent magnitude around the  RC peak. The size of the target
box is allowed  to vary in different fields, according  to the stellar density
of the field, and the number  of targets observed, spanning on average $\Delta
J_0 \approx$0.5 mag, corresponding to about 1 kpc along the the line of sight.
In Fig.~\ref{targets}  the observed stars  are color coded according  to their
metallicity (see  caption), and a color  histogram of the targets  is compared
with that  of all the RC  stars.  This is  meant to emphasize that  our target
selection  does  not  impose  any  significant  bias  in  metallicity,  except
excluding the most metal poor stars ([Fe/H]$\lsim$$-1.5$) that, if older than
10  Gyr,  would end  up  in  the blue  horizontal  branch.  The absence  of  a
metallicity bias is especially relevant in  the interpretation of the MDFs for
the two innermost field (one of which  shown here) where we find a significant
metal poor population.

Two independent sets of stars were observed  within each field, for a total of
$\sim$210 stars. The two sets are identified  as F1 and F2 in our tables.  In
a few fields  that we considered particulary interesting, four  sets of stars
were observed instead, for a total  of $\sim$430 stars.  Those fields are the
two closest to  the Galactic center, at ($0,-1$) and  ($0,-2$), and the field
at ($0,-8$) because it shows the double RC feature, signature of the presence
of     X-shape    (or     boxy/peanut    shape)     as    demonstrated     by
\citet[][]{mcwilliam+10, nataf+10, saito+11,  wegg+13}. In the ($0,-8$) field we
observed two sets of stars in the  bright RC (bRC-F1 and bRC-F2) and two sets
in the faint RC (fRC-F1 and fRC-F2).

A detailed description of the  target selection strategy, observations and the
spectrum extraction and calibration is given in Paper~I. Table~1 in that paper
lists the coordinates of all the  fields, together with the number of targets,
the spectral  setup, and the total  exposure time associated to  each of them.
Here we recall that most of the  GIBS fields were observed with the LR8 setup,
covering the Calcium II triplet  (CaT) feature (8206-9400 $\AA$) at resolution
$R=6500$.   Four  fields at  $b$$\sim$$-3.5^\circ$  where  observed at  higher
resolution using setup HR13  (6120-6405 $\AA$).   Abundances for  those are
discussed in \citet[][; hereafter Paper~II]{gonzalez+15}.

Spectra for stars  in two extra fields  were added here to  the original Large
Programme.  The  first one  consists in a  sample of 178  RC stars  in Baade's
Window,  at $(l,b)=(1,-4)$,  marked  as a  star  in Fig.~\ref{fields}.   These
stars, observed through setup HR13,  were presented in \citet[][]{hill+11} but
have been re-analyzed in Paper~II, and  we adopted here metallicities from the
latter work.  A subset (111) of these stars were also observed with setup LR8,
and 80 of them were used to derive the CaT to [Fe/H] calibration, presented in
\citet[][]{vasquez+15}  and  adopted  here.   The  second  additional  dataset
consists   of   213   RGB   stars   in  a   field   at   $(l,b)=(0,-6)$   from
\citet[][]{zoccali+08}.   Note that  these stars  are  {\it not}  the same  as
included in Paper~I.   Because it was focused on kinematics only, Paper~I  included a sample
of 454  RC stars, at  ($0,-6$), from \citet[][]{vasquez+13}, that are
more numerous than the sample analysed here.   However, those spectra  had lower S/N and,  most importantly,
the dispersion direction encompassed the  four detectors of the IMACS@Magellan
mosaic, preventing a reliable continuum definition and therefore hampering the
metallicity determination.   With the focus on  the MDF, here we  replaced the
V\'asquez dataset with the RGB stars from \citet[][]{zoccali+08}.  The latter,
although smaller in number, have been observed with the same instrument as our
HR sample.  The  analysis carried out in  \citet[][]{zoccali+08} is compatible
with the  one in Paper~II  within 0.1 dex, as  verified in the  Baade's Window
field.  The  targets at $(l,b)=(0,-6)$ are  the only ones in  the present work
that lie on the RGB, slightly above the RC.

In summary, we discuss here the MDF for 26 fields, shown in Fig.~\ref{fields}.
For  20 of  them, shown  as  red circles,  we  have LR8  spectra with  R=6500.
Metallicities were derived  by measuring CaT equivalent  widths and converting
them to  [Fe/H] following the  recipe in \citet[][]{vasquez+15}.  For  6 other
fields  (blue squares,  star and  triangle) we  have spectra  at R=22,500  and
[Fe/H] measurements were derived from equivalent widths of isolated FeI lines,
constraining also stellar surface parameters.



\section{Iron abundances}

\subsection{High resolution spectra}  \label{HR}

The analysis of  the high resolution spectra was presented  in Paper~II, where
the details  of the  adopted method  to obtain  [Fe/H]  and [Mg/Fe]  can be
found.  Here we briefly recall how [Fe/H] abundances were derived.

Stellar atmospheric  parameters were  discussed in Paper~II,  namely effective
temperature ($\rm T_{eff}$), surface  gravity ($\rm \log{g}$), microturbulent
velocity ($\rm  \xi$), and metallicity  ([Fe/H]) using the  standard iterative
method  based  on  the  equivalent  widths of  isolated  FeI  lines.   Initial
abundances  are derived  from each  line by  adopting first  guess photometric
temperature and  gravity, and these  are later refined by  imposing excitation
and ionization equilibrium.  The code GALA \citep[][]{mucciarelli+13} has been
used here to perform this task automatically.

First guess photometric temperatures were calculated by applying the calibration
by  \citet[][]{ramirez+05}  to  the   $\rm  (V-K_s)$  magnitudes  from  OGLEII
\citep{udalski+02}  and   the  VVV  catalogues  in   \citet{gonzalez+12}.  The
reddening maps in  the latter paper were also used  to correct for extinction.
Absolute V-band magnitudes were derived  using mean line-of-sight distances to
each specific  field from  \citet{gonzalez+13} and the  bolometric corrections
from \cite{alonso+99}.  These are then  used to estimate photometric gravities
based on the classical formula:

\begin{footnotesize}
\[ \rm
\log\left(g\right)=\log\left(g_{\odot}\right)+\log\left(\frac{M_*}{M_{\odot}}
\right)+0.4\left(M_{Bol,*}-M_{Bol,\odot}\right)+4\log\left(\frac{T_{eff,*}}{T_{eff,\odot}}\right)
\]
\end{footnotesize}

where    $\rm    M_{Bol,\odot}=4.72$,   $\rm    T_{eff,\odot}=5770$\,K,    and
$\log\left(g_{\odot}\right)=4.44$ dex.  As usual, we  adopted a fixed value of
$\rm M_*=0.8 M_{\odot}$.  Global metallicity and microturbulent velocity were
fixed to 0.0 and 1.5, respectively, as a first step.

An initial ATLAS9  stellar model atmosphere was constructed  using these first
guess values.  This was later updated  by GALA while iteratively searching for
spectroscopic effective temperatures and microturbulent velocity, by imposing
excitation equilibrium and the null  slope of iron abundance versus equivalent
width of the Fe line,  respectively.  As discussed in \citet[][]{gonzalez+15},
the  mean   uncertainty  in  the  metallicity   measurements  is  $\sigma_{\rm
  [Fe/H]}=0.2$ dex, reaching as low as 0.1 dex at the low metallicity end, and
up to 0.4 dex at the high metallicity end.

The  method  described  above  has  been  used,  identically,  to  derive  the
abundances for the four GIBS fields  observed at high resolution (blue squares
in   Fig.~\ref{fields})  and   for   the  Baade's   Window   giants  used   by
\citet[][]{vasquez+15} to derive the CaT to [Fe/H] calibration.  The latter is
applied  here to  the low  resolution spectra,  thus insuring  that the  final
metallicities adopted for our stars are all in the same scale.

\subsection{Low resolution spectra}  \label{LR}

\begin{figure}
\centering
    {\includegraphics[angle=0,width=1.0\columnwidth]{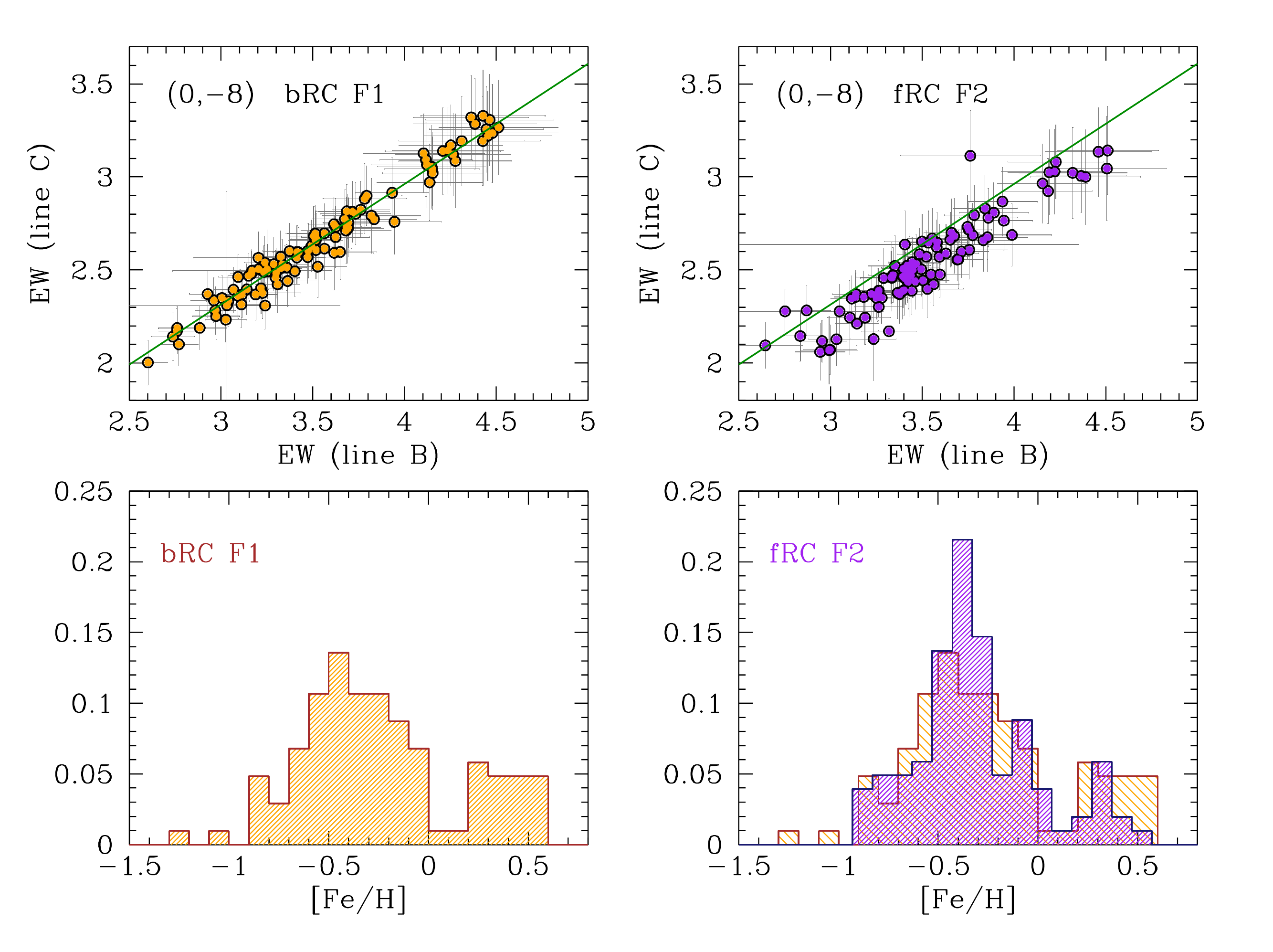}}
    \caption{Top panels:  trend of  the EW of  line C versus  line B,  for two
      different fiber allocations in the  field at ($0,-8$). The orange points
      are 103 targets in the bright RC,  while the purple ones are 102 targets
      in the faint  RC. The green line is  the same trend for the  RC stars in
      Baade's Window used to derive the  CaT vs [Fe/H] calibration.  While the
      orange targets beautifully follow the  expected ratio, the purple points
      have an offset, that might be due  either to an overestimation of line B
      and/or  to an  underestimation  of  line C.  Bottom  panels: (left)  MDF
      derived  from the  103 targets  with good  line ratio;  (right) the  MDF
      derived from the targets with offset  line ratio, in purple, is compared
      with the orange one, obtained from targets with good line ratio.}
    \label{EWs_off}
\end{figure}

The pre-reduction of  the LR8 spectra, together with the  fitting of the three
CaT  lines is  described extensively  in \cite[][]{vasquez+15}  where the  CaT
versus [Fe/H] calibration was derived. We  summarize here the main steps.  The
spectra were de-biased,  flat-fielded extracted and wave  calibrated using the
GIRAFFE pipeline  (version 2.8) provided  by ESO. Sky subtraction  was instead
applied with IRAF, by first combining the $\sim$20 sky fibres in a master sky,
which  was then  subtracted from  the  science spectra  by means  of the  task
\texttt{skytweak}.   The  spectra were  then  normalized  with the  IRAF  task
\texttt{continuum},  and finally  cross-correlation (task  \texttt{fxcor}) was
performed against a template synthetic spectrum  in order to bring the spectra
to the restframe velocity.

Equivalent widths  (EWs) of the  two strongest CaT  lines (8542 \AA\  and 8662
\AA\ ) were measured  by fitting the observed flux with the  sum of a Gaussian
component,  fitting the  line core,  plus a  Lorentzian function,  fitting the
wings.   \citet[][]{vasquez+15} presented  several tests  of the  stability of
such  fits against  metallicity and  signal to  noise, by  means of  synthetic
spectra.  They demonstrated that the weakest  CaT line at 8498 \AA\ (hereafter
line~A), does not behave smoothly with  metallicity, as the two stronger lines
(B and C), and it is the most affected by sky subtraction, due to the presence
of  two sky  lines close  to  its wings.   For this  reason the  CaT-to-[Fe/H]
calibration presented in \cite[][]{vasquez+15} relies only on the two stronger
lines.

Figure~5 in \citet[][]{vasquez+15} demonstrates that, for the stars in Baade's
Window used to derive the  CaT-to-[Fe/H] calibration, the ratio between line~C
and line~B  is remarkably constant along  metallicity.  The ratio of  these two
lines was compared with that of Baade's Window for the stars in all the fields
analysed here.   Most of the  stars indeed follow the  same line ratio  as the
calibrators in Baade's Window. In some fields, however, all the stars observed
within a  given exposure  show a  shift in the  ratio of  lineC-to-lineB, with
respect to Baade's Window (Fig.~\ref{EWs_off}).  The shift is constant for all
the  stars, i.e.,  it  does not  depend  on the  metallicity  nor the  surface
parameter of  the individual  stars. It  is also  independent from  the radial
velocity of each  star, which differs by  up to 400 km/s. We  conclude that it
must be due to an improper  continuum normalization (i.e., a residual slope in
the pseudo-continuum of  the two lines). A  variation of less than  1\% in the
EWs (upper  right panel of  Fig.\ref{EWs_off}) corresponds  to a shift  in the
pseudo-continuum  very hard  to detect.   Despite  our many  attempts, it  was
impossible to relate the presence of  the offset to any specific sky condition
(Moon, humidity, etc...),  nor to correct it.  Fortunately, the  impact of the
offset on the final metallicity is very small, if any (lower right panel).  We
therefore did  not correct for this  effect, keeping in mind  that it happens,
with the same magnitude shown in  Fig.~\ref{EWs_off}, for 10 sets of stars, in
7 fields out of the 20 observed with the CaT setup
\footnote{The affected sets of stars are the following: LRp5p4-F1, LRp5p4-F2, 
LRp0m1-F3, LRp3m2-F2, LRp0m2-F2, LRp0m2-F3, LRm8m2-F2, LRp0m8-fRC-F2, 
LRm3m8-F2, LRm8m8-F2.}.

In   order   to   apply    the   CaT-to-[Fe/H]   calibration   equation   from
\cite[][]{vasquez+15} one  needs to calculate  the so-called {\it  reduced} EW
parameter, $W^\prime$, defined as
\[
W^{\prime}=\Sigma \mathrm{EW} + 0.384 (K-K_{\rm RC})
\]
where $\Sigma \mathrm{EW}$ is the sum of the EWs of line~B and line~C, and the
term  $0.384  (K-K_{\rm  RC})$,  is  a  correction  that  takes  into  account
simultaneously the effect of the surface  gravity and temperature of the star,
parametrized as the magnitude difference between  the star and the RC, in that
specific field.

To that end,  the observed $K_s$ magnitudes of all  stars were corrected using
the interstellar extinction maps  of \cite[][]{gonzalez+12} before calculating
the  difference  with  respect  to   peak  dereddened  RC  magnitude  in  that
field. Since the  stars were chosen to  be on the RC, this  difference is very
close to zero for all our  targets.  We nevertheless included this correction,
because the target  selection was done on the observed  magnitudes, before the
extinction map  was published.   Therefore, in  the innermost,  most extincted
fields, some  stars might end  up slightly displaced from  the peak RC  in the
de-reddened CMD. The CaT calibration presented in \citet[][]{vasquez+15} has a
rms uncertainty of 0.2 dex.

\begin{table*}
\caption{Coordinates, VVV magnitudes, radial velocities and metallicities of the program stars.}
\label{metallicities}
\centering
{\small
\begin{tabular}{c c c c c r c c}
\hline\hline
ID & RA (J2000) & DEC (J2000) & $J$ & $K_s$ & RV & Err RV & [Fe/H] \\
\hline
   LRp0m1\_F1\_OGLE5\_140424 &  17:50:20.66 &  $-$29:58:42.30 & 14.617 & 13.234 &  $-$42.9 &  2.1 &  $-$0.33 \\
   LRp0m1\_F1\_OGLE5\_141397 &  17:50:19.03 &  $-$29:58:04.10 & 14.761 & 13.361 &  198.0 &  2.3 &  $-$0.65 \\
   LRp0m1\_F1\_OGLE5\_145942 &  17:50:19.62 &  $-$29:56:20.40 & 14.166 & 12.551 &  253.0 &  2.4 &  $-$0.10 \\
... & ... & ... & ... & ... & ... & ... & ...  \\
\hline
\hline\hline
\end{tabular}}
\end{table*}

The coordinates, magnitudes,  metallicities and radial velocities  of the GIBS
targets   discussed   here   are   given  in   the   electronic   version   of
Table~\ref{metallicities}.
	
\section{The bulge Metallicity Distribution Function}

\begin{figure*}[ht]
\centering
    {\includegraphics[angle=-90,width=17cm]{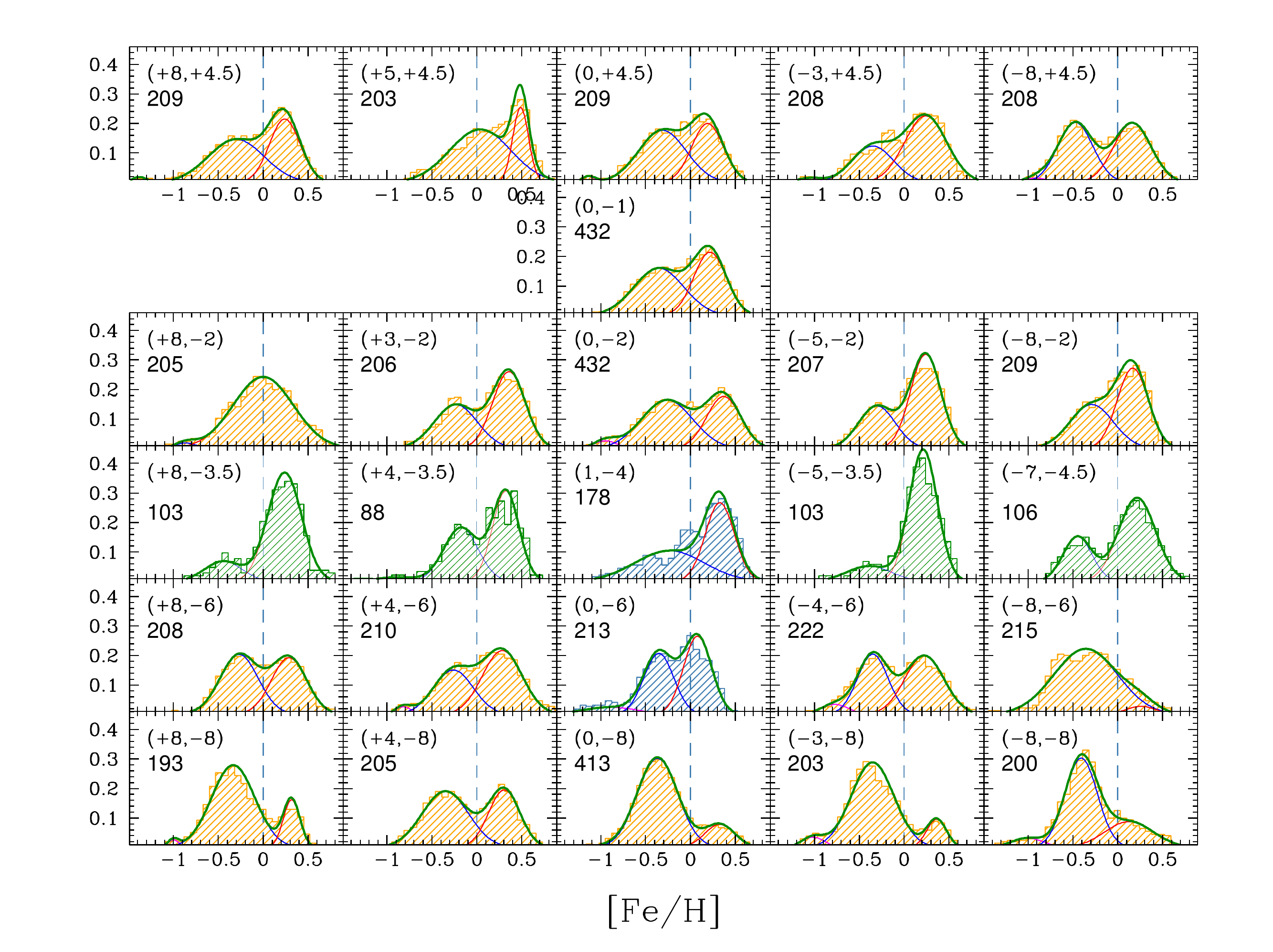}}
    \caption{Metallicity distribution function (MDF) for  the 26 fields of the
      present study.  Orange histograms  have been  derived from  CaT spectra,
      while green histograms  come from the high  resolution spectra discussed
      in Paper~II.  Blue histograms are  from our previous works.   Panels are
      arranged  like  the fields  in  the  sky, with  approximate  coordinates
      indicated in the  labels. The number of stars observed  in each field is
      also labelled for each field.}
    \label{MDFs_all}
\end{figure*}

\begin{figure}[h]
\centering
\includegraphics[angle=0,width=9cm]{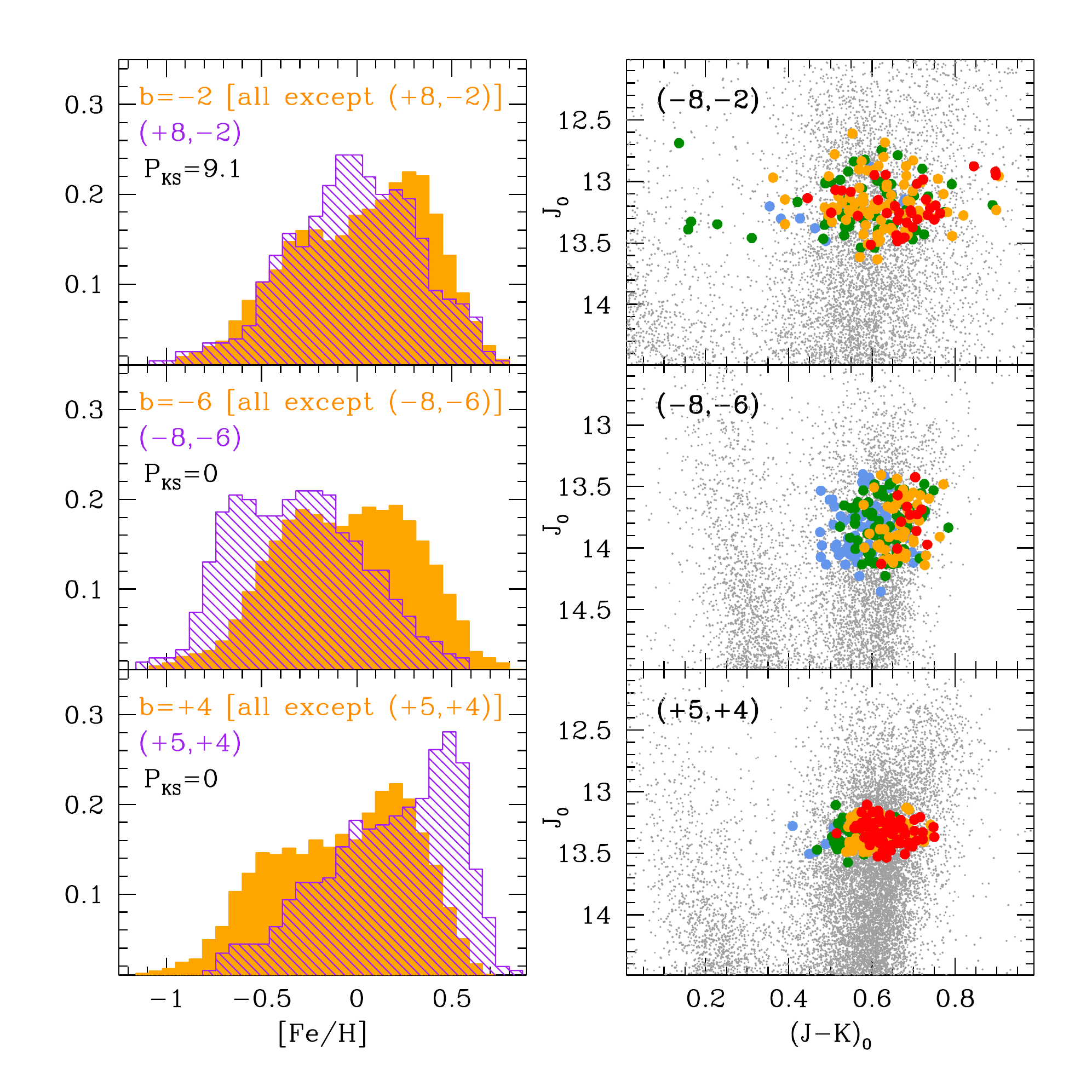}
\caption{Left panels: MDF  of the discrepant fields, compared  with the global
  MDF  of  the  other  fields  at  the same  latitude.   The  p-value  of  the
  Kolmogorov-Smirnov test comparing  these peculiar MDFs to the  global MDF is
  indicated.  Right  panels: corresponding target selection  for these fields.
  Targets  are color-coded  according  to their  metallicity,  with blue  dots
  having    [Fe/H]$<-0.5$    dex,     green    $-0.5<$[Fe/H]$<-0.1$,    orange
  $-0.1<$[Fe/H]$<+0.3$ and red [Fe/H]$>+0.3$. }
\label{make_ks}
\end{figure}

The resulting metallicity  (iron) distribution function for the  26 fields is
shown in Fig.~\ref{MDFs_all}.  Panels in this figure are arranged in the same
way as the fields in Fig.~\ref{fields}, with labels within each panel showing
the galactic coordinates  of each field, and the number  of stars included in
the histogram.  The orange histograms  refer to  the fields observed  at low
resolution (Sec.~\ref{LR}),  the green histograms are the fields  observed at
high resolution  (Sec.~\ref{HR}) with our  ESO Large Programme, and  the blue
histogram are  the 213 RGB  stars from \citet[][]{zoccali+08} at  (0,-6), and
the 178 RC stars from \citet[][]{hill+11} in Baade's Window.

All the  histograms presented in  this Section  have been obtained  as Average
Shifted  Histograms.  Specifically,  three histograms  were constructed,  each
with a bin of 0.21 dex  and starting points at [Fe/H]=$-2.0,-1.93$ and $-1.86$
(i.e., shifted by 0.07 dex), respectively. A vertical line, fixed at [Fe/H]=0,
has  been drawn  in each  panel, as a guide to the eye.   All the  fields in
Fig.~\ref{MDFs_all} show a  clear bimodality, except for two  of them (LRp8m2,
LRm8m6)  that look  unimodal.   In order  to assess  the  significance of  the
bimodality,  we  have  run  a Gaussian  Mixture  Model  \citep[GMM][]{muratov+10}
allowing for the presence of one,  two, or three Gaussians with different mean
and dispersion.   The best fitting  solution, for most  of the fields,  is the
trivariate one, with a tiny peak for [Fe/H]$<-0.8$.  This peak is usually made
up of a handful of  stars, and in some fields is so small  that it is not even
visible in Fig.~\ref{MDFs_all}.   Suppressing this peak, and  fitting only two
Gaussians, however, resulted the metal  poor Gaussian being very broad, enough
to  include   these  few   stars,  and   therefore  extending   to  supersolar
metallicities. This peak would unrealistically include some of the most metal
rich  stars of  the  sample. The triple Gaussian fit did not converge in four 
fields (LRp5p4, LRm5m2,  HRp8m3, Baade's Window); for these fields the data
were best fit with double Gaussians.

\begin{figure}[h]
\centering
\includegraphics[angle=0,width=8cm, height=8cm]{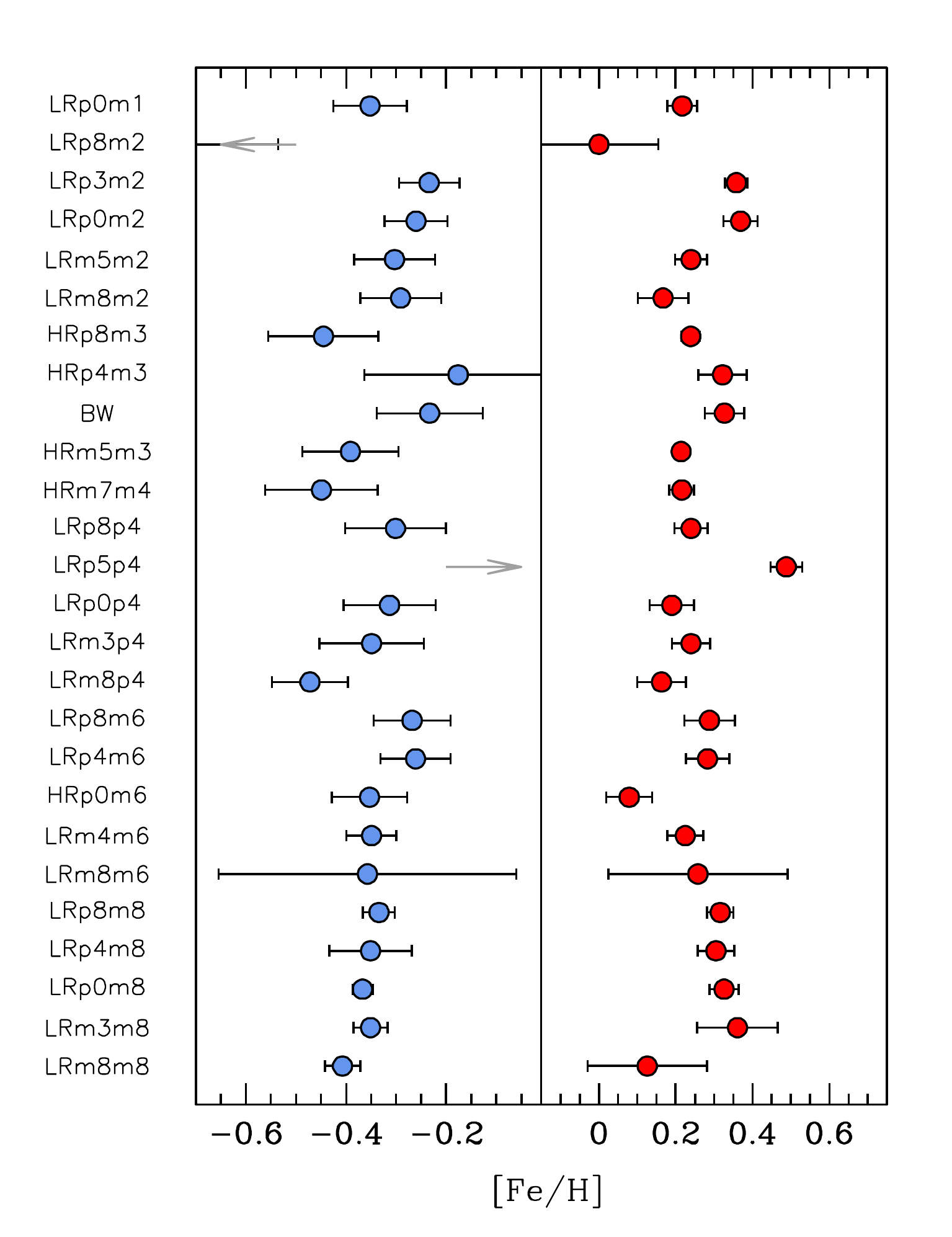}
\caption{Peak metallicity of the metal poor (blue) and metal rich (red) 
Gaussians fitted in Fig.~\ref{MDFs_all}. The name of the fields are listed on 
the left, from the closest to the plane at the top, to the ones in the outer
bulge at the bottom.}
\label{peaks}
\end{figure}

In  fields  LRp8m2  and  LRm8m6  the  two  Gaussian  model  did  not  yield  a
significanly  better  fit  compared  to  a single  Gaussian.   It  would  seem
unrealistic  that  the  bulge  contains two  metallicity  components,  with  a
separation of  $\sim$0.6 dex in  [Fe/H], {\it except}  in these two  fields.  We
therefore double checked  the target selection and  the velocity distributions
of  stars in  these  two fields,  to  ensure  that we  are  sampling the  same
populations as  in the others.  From  Fig.~\ref{feh_rv} below, we see  that at
($+8,-2$) and ($-8,-6$), the velocity  versus metallicity scatter plots do not
show   any   strange   feature, that could indicate a possible sub-population.

Fig.~\ref{make_ks} (left panels) shows  the p-value of Kolmogorov-Smirnov test
comparing the MDF  of these fields to  the global MDF for the  other fields at
the same  latitude. In the  case of the  ($+8,-2$) field, the  p-value ($9\%$)
does not permit rejection of the hypothesis  that this field is drawn from the
same MDF.
The  top right  panel shows  the target  selection for  this field:  while the
selection was  done on the  observed CMD (before  the VVV reddening  maps were
available), de-reddened magnitudes  are shown here; therefore, the targets are
no longer confined to a CMD box. This field in particular  is very close to
the plane, where differential reddening  is non negligible.  After correction,
some targets seem to  belong to the foreground disk rather  than to the bulge.
Targets here are  color coded according to their metallicity:  blue points are
stars with [Fe/H]$<-0.6$ dex, green with $-0.6<$[Fe/H]$<-0.1$ dex, orange with
$-0.1<$[Fe/H]$<+0.3$ dex,  and red with [Fe/H]$>+0.3$  dex. Contaminating disk
stars all belong to the peak at [Fe/H]$\sim -0.2$ dex.  Indeed, removing stars
with color  $(J-K)_0<-0.35$ would  change the p-value  to 15$\%$.  We conclude
that this  fields happened  to have  a larger disk  contamination, due  to the
higher differential reddening.  Its  distribution is anyway compatible, within
the statistical  fluctuations of the  present sample,  with the global  one at
this latitude. Note also that the metallicity derived for disk stars would not
be  accurate,  since their  gravity  differ  significantly  from that  of  CaT
calibrating stars.

Stars in the  field at ($-8,-6$) are  shown in the middle panels.  Here the MDF is
much more metal  poor than the global one, with  null probability ($10^{-11}$)
of  having been  drawn from  it.  On  the other  hand, the  CMD does  not show
obvious contaminations, and neither does  the velocity versus metallicity plot
in Fig.~\ref{feh_rv}. We do not have  a plausible explanation for the observed
difference  between  the  MDF  of  this  field and  the  global  one  at  this
latitude. Even if a unimodal Gaussian distribution would fit the MDFs of these
two fields marginally better than a bimodal  one, we imposed a bimodal fit for
consistency with  the other fields. Obviously  the position of the  peaks will
not be well constrained in this case.

Finally, another rather  peculiar field is the one at  ($+5,+4$). In this case
the MDF has two main peaks like all the others, but it is all shifted to metal
rich values.  Again,  the target selection does not seem  to be different from
that of the other  fields. This is one of the fields were  the ratio of line B
to line  C has  an offset  with respect  to the  calibrating stars  in Baade's
Window.  We have shown in Fig.~\ref{EWs_off} that the impact of this offset on
the MDF  is negligible, but  in this  field we have  an offset in  both target
datasets (LRp5p4-F1 and LRp5p4-F2) hence we cannot really verify that. We will
treat this field like the others, keeping in mind its peculiarity and ensuring
that none of the conclusions drawn in the present paper is based on any of the
fields in Fig.~\ref{make_ks}.  Rather, we will be able to  draw robust results
{\it in spite of} the peculiar shape of the MDF in those fields.

In what  follows we will try  to characterize the two  metallicity populations
that  we  see   in  the  bulge.  We   will  assume  that  the   tiny  peak  at
[Fe/H]$\sim$$-1$ is made up by contaminating halo stars.  Notice that our MDFs
do not  include very metal poor  stars because our target  selection criterion
excludes stars outside the color range  of the RC.  Stars with [Fe/H]$<-1$, if
they are old enough,  would end up to the blue of the  RC, and would therefore
fall  outside  the  target  selection  box.    The  MDFs  for  the  fields  at
$|b|>3^\circ$ confirm  previous findings by our  group \citep[][]{zoccali+08},
the   ARGOS   survey   \citep[][]{ness+13mdf}    and   the   Gaia-ESO   survey
\citep[][]{rojas-arriagada+14}.  Namely,  that the  bulge has  two metallicity
components: with the metal poor one dominating at high latitude, and the metal
rich dominating closer  to the plane.  The mean metallicity,  therefore, has a
gradient which is due  to the different mix of the two  populations, and it is
mostly vertical,  rather than  radial (Paper~II).  Fig.~\ref{peaks}  shows the
metallicity of  the metal  poor (blue)  and metal rich  (red) peaks  as fitted
above.  Errobars  were  calculated  by means  of  a  non-parametric  bootstrap
(drawing  from  the  input  sample   with  repetitions)  of  100  realizations
\cite[][]{muratov+10}.   Overall, the  metal poor  and metal  rich peaks  stay
constant, within the  errors, with only two fields rather  discrepant, both of
them discussed above.

\begin{figure}[h]
\centering
    {\includegraphics[angle=0,width=9cm, height=10cm]{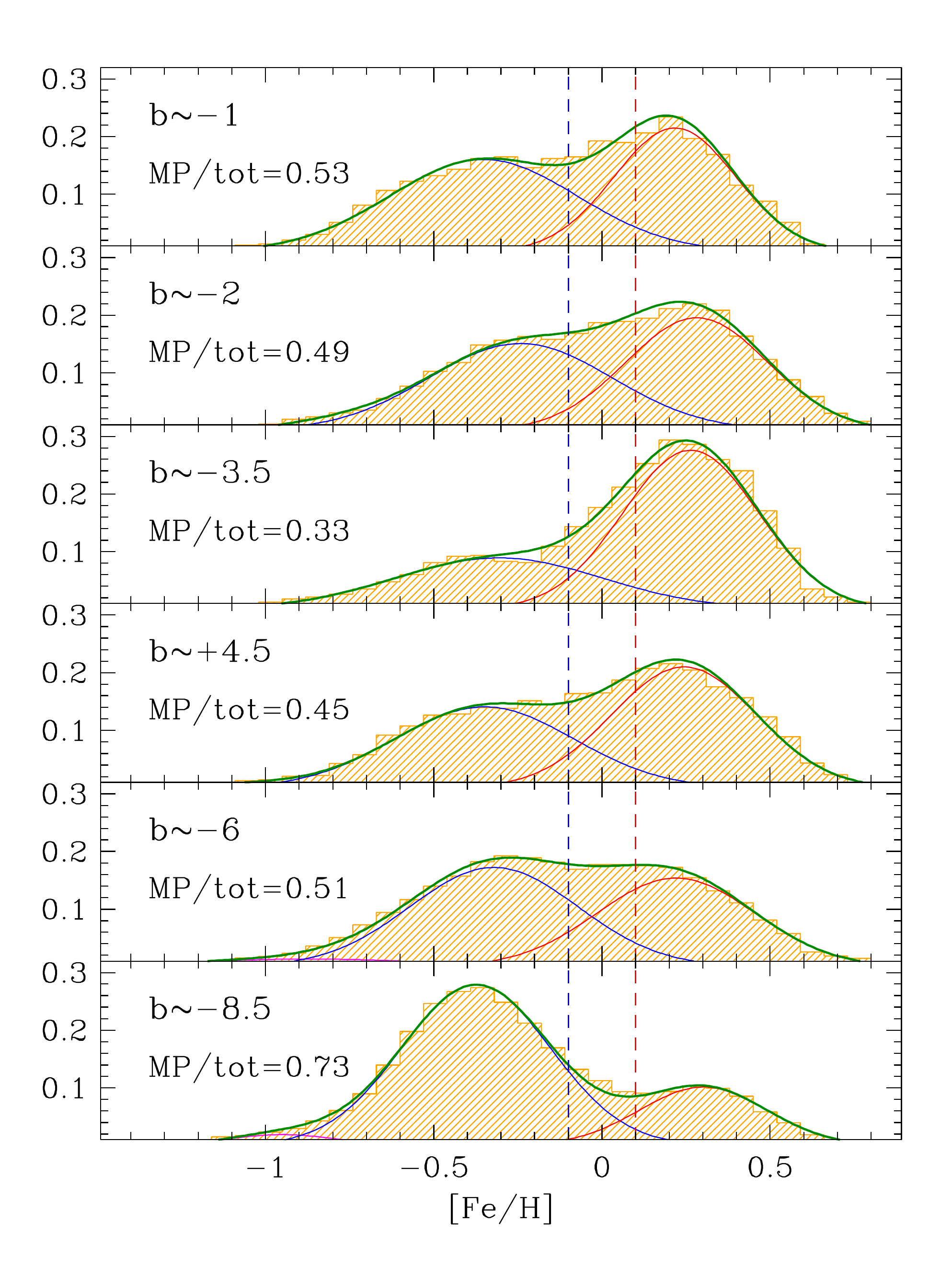}}
    \caption{MDFs of all the stars at constant latitudes, where fields
    at different longitudes have been combined. The fraction of metal
    poor stars, compared to the total, is listed in each panel. As
    discussed in the text, the fraction of metal poor stars decreases
    from $b=-8.5$ to $b=-3.5$, where it reaches a minimum, but then it
    rises again at latitude closer to the Galactic plane. The vertical
    dashed lines mark the limits of the metal poor and metal rich populations
    (see text). }
    \label{MDFs_bcost}
\end{figure}

The most important  result from the present data is  that, at latitude smaller
than $|b|=3^\circ$, the metal poor  component becomes important again. Because
the variation of the MDF with longitude, at constant latitude, is smaller than
the variation  as a  function of height  from the plane,  we have  coadded the
stars  in  all the  fields  at  constant latitude,  and  show  the results  in
Fig.~\ref{MDFs_bcost}.  Thanks  to the  larger statistics,  the MDFs  here are
much smoother and the two Gaussian  fits to stars with [Fe/H]$>-0.8$, obtained
with the GMM, are  excellent.  The label in each panel  states the fraction of
metal poor stars, i.e., the relative  contribution of the blue Gaussian to the
total  bimodal   fit.   This  fraction   is  0.77  at   $b=-8.5^\circ$,  drops
continuously when  moving closer to the  plane, reaching a minimum  of 0.29 at
$b=-3.5^\circ$,  and then  goes back  up to  0.6 at  $b=-2^\circ$ and  0.53 at
$b=-1^\circ$.

The MDF for the innermost $\sim$ 2 degree from the Galactic center was derived
by \citet[][]{rich+07}  and \citet[][]{rich+12}.  They analysed near  IR, high
resolution spectra for  $\sim$ 15 M giants  in each of two  fields at ($0,-1$)
and ($0,-1.75$), respectively, and found only metal  rich stars, with only a few
of    them    having    [Fe/H]$<-0.3$,   and    none    with    [Fe/H]$<-0.5$.
\citet[][]{babusiaux+14}, on the other hand,  derived the MDF from CaT spectra
for 107 giants in  a field at ($0,+1$). They do find  a metal poor population,
rather similar  to the present one  at ($0,-1$), although with  a longer metal
poor  tail, possibly  due  to  the slightly  different  target selection  box.
Finally, the presence  of a metal poor  component within a radius  of 1 degree
around the Galactic  center was detected by  \citet[][]{schultheis+15}, from a
sample of 33 M giants observed within the APOGEE data.

Two things are very clear from Fig.~\ref{MDFs_all} and Fig.~\ref{MDFs_bcost}:
{\it i)} the bulge does have two metallicity components; and {\it ii)} the
spatial distribution of the components is different. In order to investigate the
density distribution of the two components we need to convolve the information
given in Fig.~\ref{MDFs_all} with what we know about the bulge global density
distribution. In other words, although the metal poor component is dominant
in the outer bulge (at $b=-8.5^\circ$) it might not be very conspicuous, just
because the stellar density at these latitudes is rather low. We investigate
this specific point in the next Section.

\section{Characterizing the two metallicity populations in the bulge}

In this Section  we will try to characterize the  spatial distribution of the
metal poor  and metal rich stars,  separately.  We will also  investigate the
radial  velocity distribution  of the  two populations.  Let us  stress, that
although  it  will  be  clear  that  these  two  populations  have  different
properties, this does  not necessarily imply that they  had different origin.
Certainly, the difference  in spatial and kinematical  properties between the
two metallicity  components must  be taken into  account when  studying bulge
formation models  which should  ultimately provide the  link to  the physical
processes behind them. The purpose of  this paper, however, is to present and
describe the observational data only.

\subsection{Density distribution of red clump stars}

\begin{figure}[h]
\centering
\includegraphics[angle=0,width=8cm]{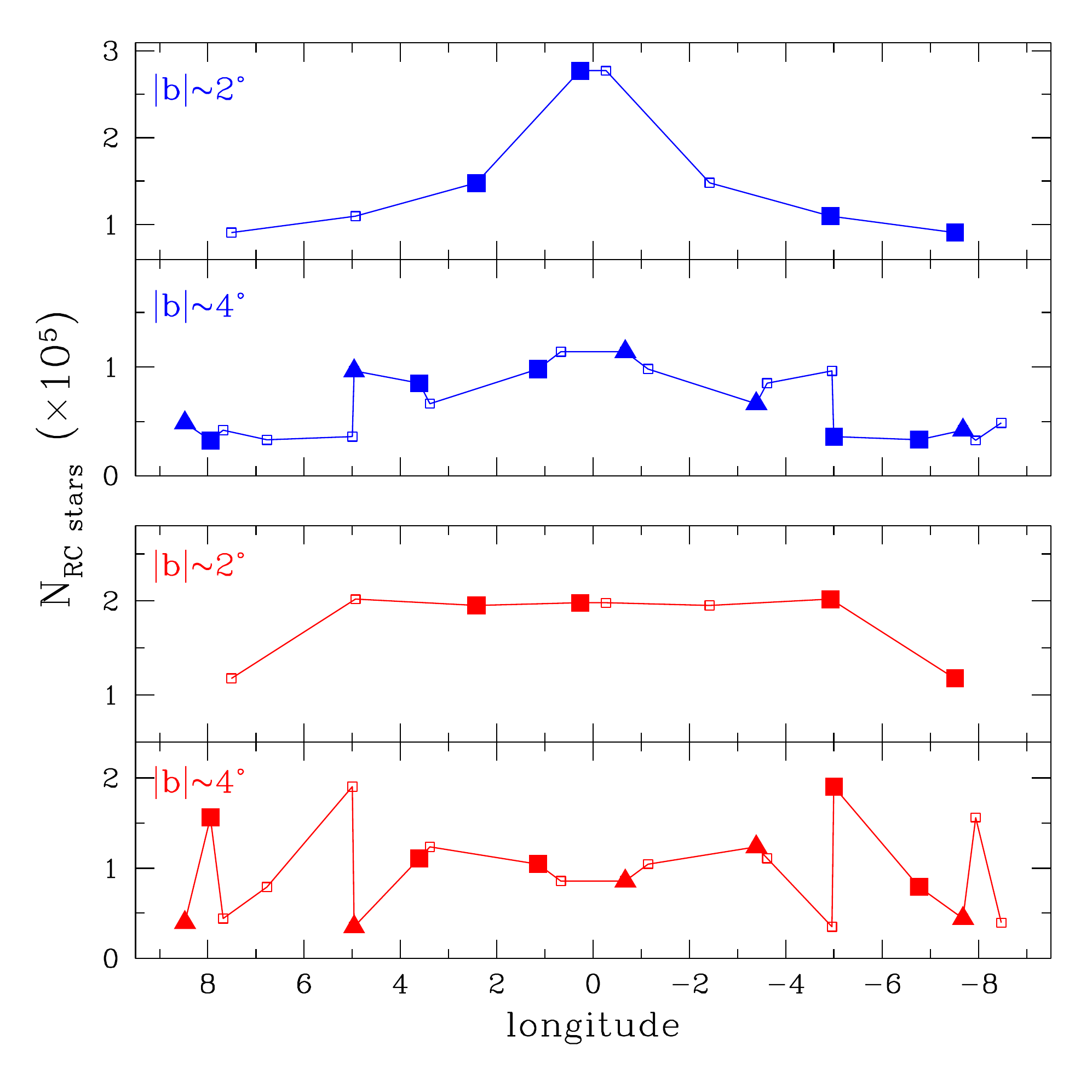}
\caption{Stellar density of RC stars, assumed to trace the global stellar
density, for the metal poor (top panels, blue) and the metal rich (bottom
panels, red) components. Filled squares/triangles refer to actual fields 
at negative/positive latitude, respectively. Open symbols are their mirror
positions, when folded in longitude. Two strips at almost constant latitude
are shown here, as indicated in the labels. Notice that the metal poor component
has a steeper gradient towards the center, while the metal rich one has a
very mild, if any, gradient at these latitudes.}
\label{fractions}
\end{figure}

Using  the  fact that  the  bulge  MDF can  be  parametrised  by two  Gaussian
components, each  having a coherent  spatial variation, the fraction  of metal
poor to metal rich stars given by the MDF in each field can be translated to a
stellar density of each component individually.  This can be done by using the
bulge stellar density map from  \citet[][]{valenti+16}, who counted the number
of RC stars across the bulge using PSF photometry from the VVV survey.  We use
this number  and the relative  fraction of metal-poor and  metal-rich Gaussian
components of each of the GIBS fields to calculate the corresponding number of
metal poor and metal rich RC stars, in each field.  The number of RC stars for
each component of all fields is listed in Table~\ref{table_mdfs}.

In order  to visualize the  differences in  the spatial distribution  of metal
poor  and  metal rich  RC  stars,  Table~\ref{table_mdfs}  should be  used  to
construct spatial  density maps  for the two  components. This  is challenging
with the  current data, because the  GIBS field grid is  relatively coarse. In
order to  improve the  spatial resolution,  we folded the  field grid  both in
Galactic longitude  and latitude.   When doing  so, we  are imposing  that the
density maps  must be symmetric both  about $l$ and $b$.  While symmetry about
the Galactic  plane is very plausible,  the orientation angle of  the Galactic
bar does  produce an asymmetry in  the projected density between  positive and
negative  longitudes   \citep[][]{alard+01,  valenti+16}.   For   the  present
investigation, however, we are assuming that  our data are not dense enough to
allow us to detect differences between the near and far sides of the bar.

Figure~\ref{fractions} shows the  number of RC stars in  two strips at
approximately constant latitude,  as a function of  longitude, for the
metal poor (top panels, in blue) and the metal rich (bottom panels, in
red) components. Filled symbols show  the actual positions of the GIBS
fields, at negative (squares) and positive (triangles) latitude.  Open
symbols  are the  mirror positions  of  these fields,  when folded  in
longitude.  The two  discrepant  fields LRp8m2  and  LRm8m6 have  been
omitted in this analysis.  Figure~\ref{fractions} shows that the metal
poor component is peaked towards the  center, while the metal rich one
has  a  fairly  constant  density  distribution,  at  least  at  these
latitudes.   The  rather flat  distribution of  metal rich  stars
  resemble the RC surface density in \citet[][their Fig.13]{Nataf+15},
  at $|b|$=5.5,  although the  latter study included  all the  RC stars,
  without distinction on metallicity. Figure~\ref{fractions} shows a
subsample of  the real data,  without any modeling  nor fitting.
It has however one important limitation, because the fields shown here
have  only  {\it  approximately}  constant  latitude:  the  fields  at
$|b|\sim$4$^\circ$  actually  span  a range  from  $|b|=3.5^\circ$  to
$|b|=4.5^\circ$.  Other fields not shown  here are even more scattered
at different  coordinates.  A deeper  analysis of what the  data show,
therefore,   requires  an   interpolation   and   a  two   dimensional
visualization.

\begin{figure}[h]
\centering
\includegraphics[angle=0,width=9cm]{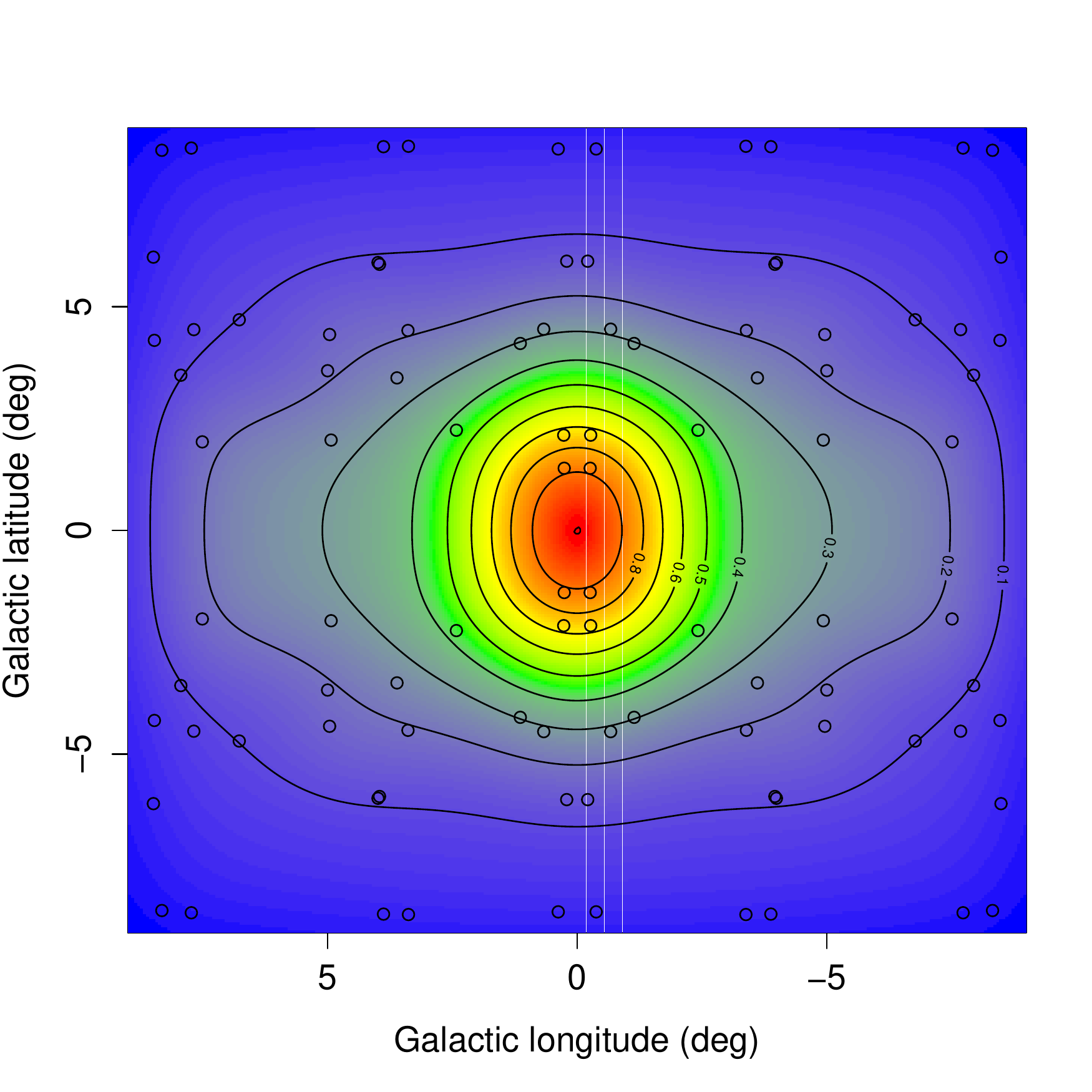}\\
\includegraphics[angle=0,width=9cm]{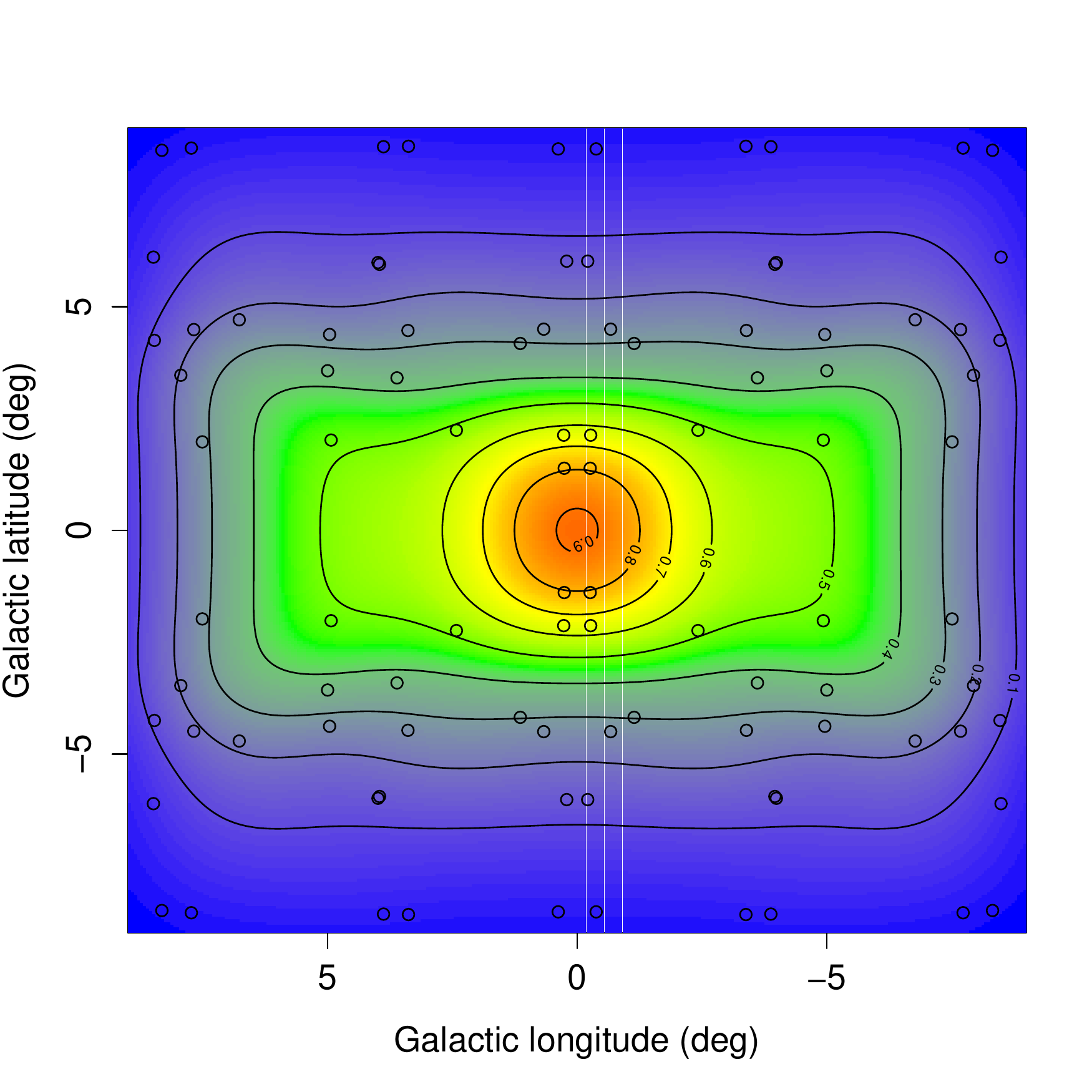}
\caption{Density map  of metal-poor (top)  and metal-rich (bottom)  RC stars
  obtained using the two component MDF in  each field and the total number of
  RC stars from \citet[][]{valenti+16}.  Open circles show the grid of fields
  used in the interpolation, resulting from  folding the original map of GIBS
  fields with respect to both Galactic  latitude and longitude. The number of
  RC stars of each  component were normalised by the maximum  of number of RC
  stars  in the  metal-poor map  so they  are color-coded  to the  same scale
  between  values of  0 and  1.  Contours are  overplotted in  both maps  and
  correspond to differences of 0.1 in the normalised RC star counts.}
\label{metal-density}
\end{figure}

We applied  a bivariate  linear interpolation  to the  irregular grid  of GIBS
fields using the  \textit{akima} package in R (Akima 1978)  to produce stellar
density maps  of the  metal poor and  the metal rich  components, as  shown in
Fig.~\ref{metal-density}.  To construct the maps,  we normalised the number of
RC stars  of each component  by the maximum number  of metal-poor RC  stars so
that the  color code of the  maps can be  compared directly.  We focus  on the
general shape  of the  maps instead of  the details as  the latter  are highly
dependent  on the  interpolation method  and the  non-uniform grid.   For this
reason, we smoothed the  map using an  isotropic Gaussian
kernel with  standard deviation of  1 deg. The  fields LRp8m2 and  LRm8m6 have
been omitted from  the maps because the  fraction of metal poor  to metal rich
component obtained  from them are  not reliable in  our opinion. In  any case,
thanks to the  four fold symmetry that  we chose to impose,  the maps obtained
with  these fields  included are  only  barely distinguishable  from the  ones
presented here.

It is clear from Figure~\ref{metal-density} that the density map of metal-poor
RC  stars  is  more  concentrated  than  that  of  the  metal-rich  RC  stars.
Furthermore,  the  metal-rich  map  appears more  elongated,  or  boxy,  along
Galactic longitude because  of a steep drop in the  stellar counts after |b|=5
at all Galactic  longitudes. On the other hand, towards  the central region of
the  maps,  the metal-poor  stars  are  not only  seen  to  be more  centrally
concentrated but they also overcome the number of metal-rich RC stars.

\subsection{Kinematics versus Metallicity trends}

\begin{figure*}
\centering
    {\includegraphics[angle=-90,width=18cm]{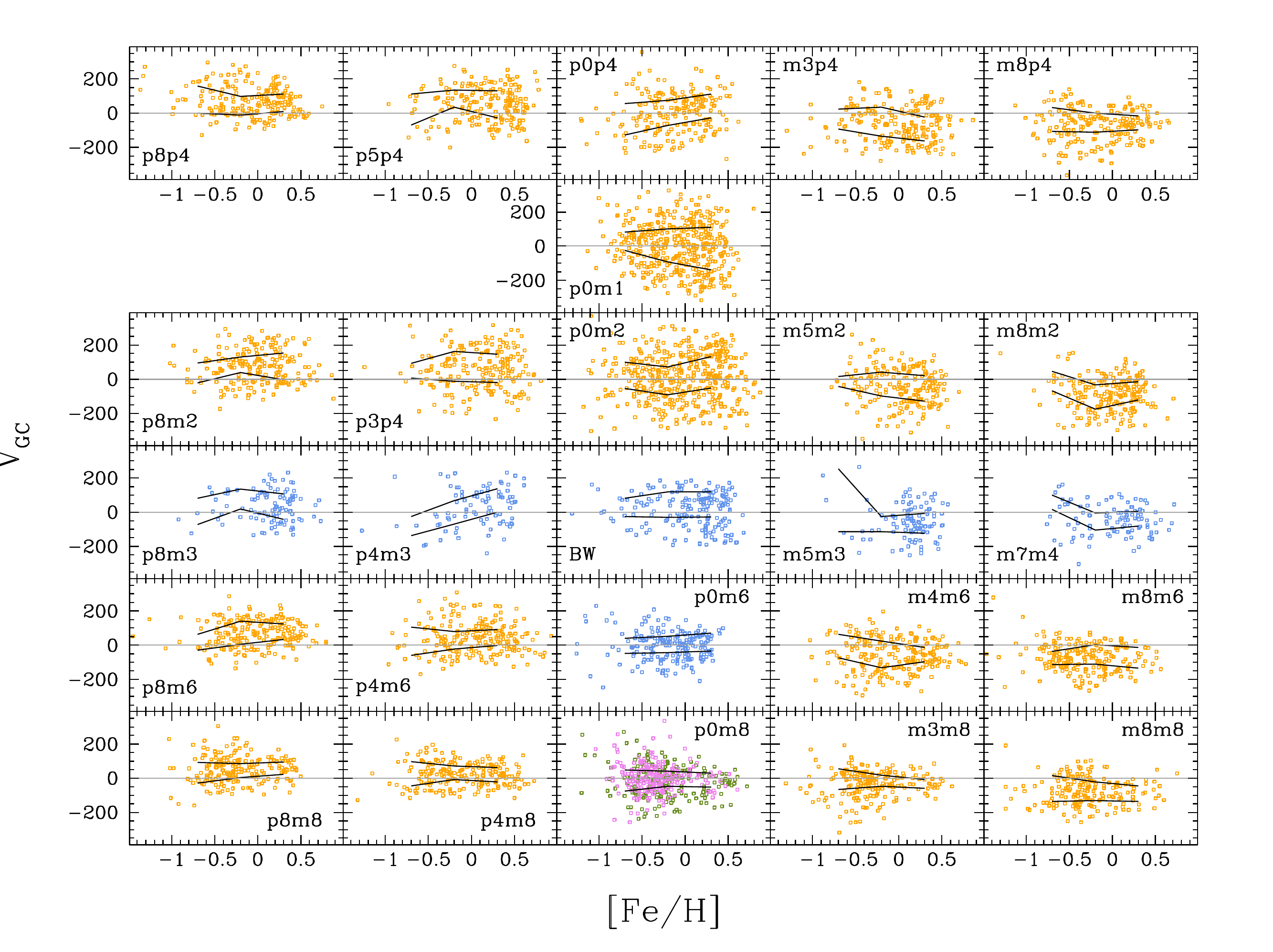}}
    \caption{Metallicity  versus galactocentric  radial velocity  for all  the
      targets in  the GIBS fields.  Orange  points are stars observed  in CaT,
      azure  points  are stars  observed  at  high  resolution. The  field  at
      ($0,-8$) has  a double RC:  bright-RC targets  are shown in  green while
      faint-RC ones are shown in pink. The solid lines mark the $\pm 1 \sigma$
      interval around the mean velocity, at three different metallicities. See
      text for details.}
    \label{feh_rv}
\end{figure*}

We  present here  a kinematical  characterization  of the  sample stars,  with
particular emphasis on  possible differences between the metal  poor and metal
rich  components.   Only  the  radial velocities,  and  velocity  dispersions,
available from the GIBS spectra are discussed here. The proper motions for the
spectroscopic targets are  currently under analysis from VVV  data.  They will
allow  us  to calculate  three-dimensional  space  velocities, which  will  be
discussed in a forthcoming paper.

Figure~\ref{feh_rv}  shows the  galactocentric radial  velocity of  the sample
stars versus  their [Fe/H], for all  the fields. Orange points  are the fields
observed with the LR8 setup, in the CaT spectral region. Light blue points are
the targets observed  at higher spectral resolution. Green and  pink points in
the  field  at  $(0,-8.5)$,  refer  to  stars in  the  bright  and  faint  RC,
respectively, which were also observed with  the LR8 setup.  This figure shows
that  the distributions  are fairly  homogeneous.   Although there  are a  few
outliers with respect  to the main radial velocity distribution,  there are no
obvious bimodalities, nor tight clumps suggestive of possible streams.  Due to
the bulge rotation, stars in the  fields at positive longitudes (panels on the
left  side of  this figure)  have positive  (receding) mean  velocities, while
stars  in  fields at  negative  longitudes  have negative  (approaching)  mean
velocities.

\begin{figure}[h]
\centering
    {\includegraphics[angle=0,width=9.5cm]{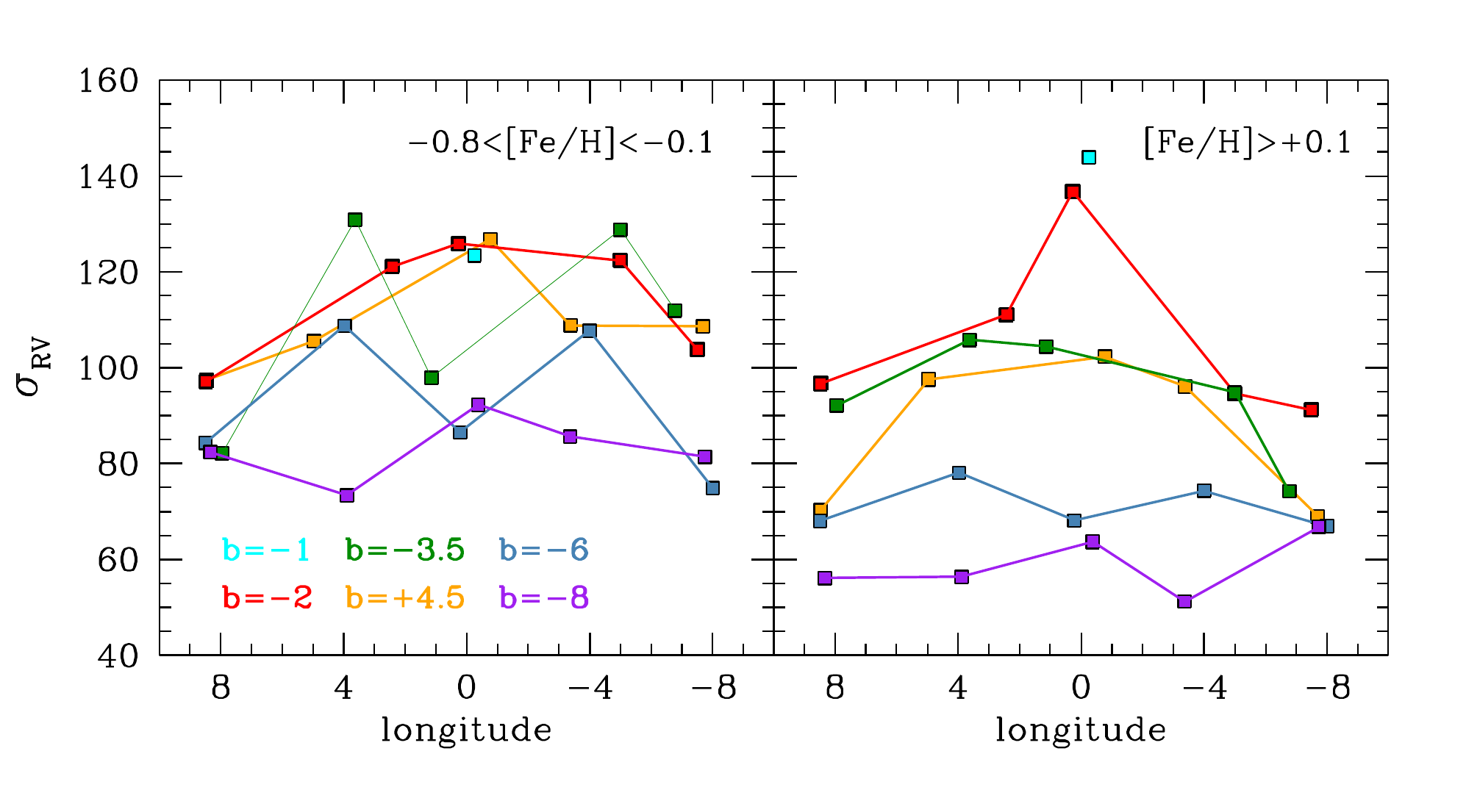}}
    \caption{Velocity dispersion as a function of longitude, for fixed latitudes
    shown in different colors. Metal poor stars are shown in the left panel,
    metal rich ones in the right panel.}
    \label{sig}
\end{figure}

This  figure does  not show  any obvious  trend of  the mean  radial velocity
versus metallicity.  It does show,  however, that the velocity dispersion has
a trend as a function of  metallicity.  Comparing, for example, the innermost
field (second row from the top) with the outer one along the minor axis (pink
and green  points) the velocity dispersion  of the metal poor  stars, smaller
than that of the  metal rich stars in the innermost  field, becames larger in
the outer field. This change is more  evident along the minor axis, but is is
roughly conserved at $l\neq0$.  The black broken lines within each panel show
the $\pm 1\sigma$  interval around the median, as a  function of metallicity,
for  three 0.5  dex  bins centered  at $[Fe/H]=-0.7$,  $-0.2$  , and  $+0.3$,
respectively,  with the  aim  of  emphasizing the  change  of behaviour.   As
expected, the  stars observed at  high resolution at $b\sim-3.5$  (light blue
points) show a very noisy behaviour, when binned in metallicity, because they
are  fewer in  number,  and also  because  at this  latitude  the metal  poor
component is very small compared to the metal rich one.

\begin{figure}
\centering
    {\includegraphics[angle=0,width=1.0\columnwidth]{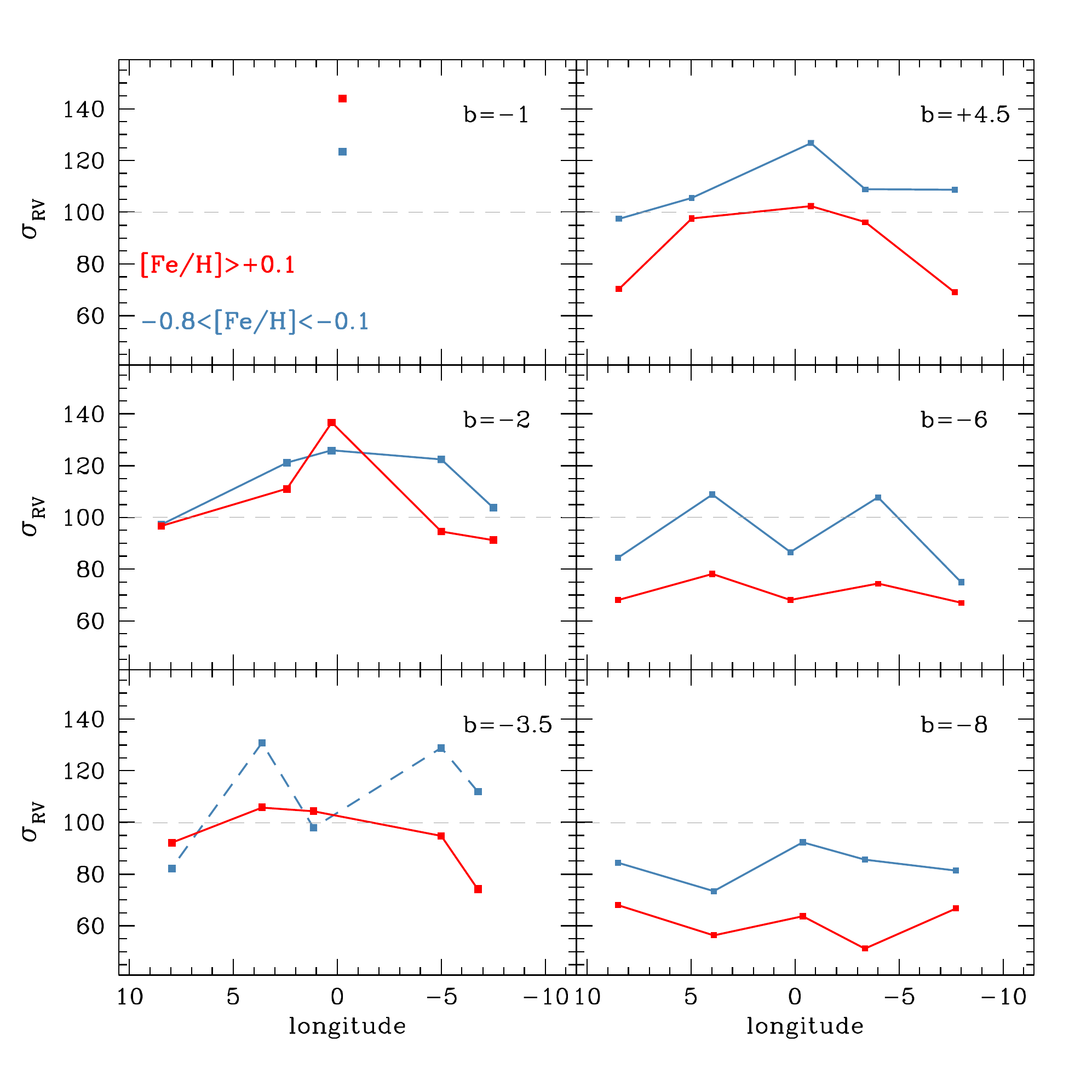}}
    \caption{The trend of  velocity dispersion with longitude, in  6 strips of
      constant latitude.  The horizontal  dashed line at $\sigma_{\rm RV}$=100
      has been drawn to guide the eye.  Notice that metal poor component has a
      higher velocity dispersion, compared to the metal rich one, in the outer
      bulge $b>3.5$.  However at $b=-2$ $\sigma_{\rm RV}$ is roughly equal for
      both metallicity bins, while at  $b=-1$ the $\sigma_{\rm RV}$ for the metal
      poor sample  lies below that of  the metal rich population.  Notice that
      the metal poor  component at $b=-3.5$ contains very few  stars, for this
      reason the trend is noisy, it is shown with a with dashed line.}
    \label{sig_bcost}
\end{figure}

The variation of  the velocity dispersion discussed above  is best illustrated
in Fig.~\ref{sig} and Fig.~\ref{sig_bcost}, showing the velocity dispersion as
a function  of longitude, for  different latitude,  splitting the sample  in a
metal poor and a metal rich one. With the aim of separating the two components
we have decided  to make a constant  cut at [Fe/H]$=0.0$, with  the metal poor
sample composed  by stars with  [Fe/H]$<-0.1$ dex,  while the metal  rich only
contains   stars    with   [Fe/H]$>+0.1$    dex.    The   narrow    strip   at
$-0.1<$[Fe/H]$<+0.1$ has been  left out as a {\it no  man's land}, because the
large cross contamination would add more noise than number statistics. We have
shown in  Fig.~\ref{peaks} that the peaks  do not move significantly  from one
field   to  the   other,  hence   the  choice   of  a   constant  cut   (e.g.,
Fig.~\ref{MDFs_bcost}). The GMM yields the membership probability of each star
to either  population. However, in the  overlap region, no model  can possibly
establish the membership probability on a one by one star basis. Therefore, we
rather  exclude  the   region  where  the  overlap   between  the  metallicity
distribution is significant.

Figure~\ref{sig}  confirms   what  was  already   shown  in  the   ARGOS  data
\citep[][]{ness+13mdf}, i.e., that the sigma of metal poor stars is on average
higher than that of metal rich stars, at least for $|b|>4^\circ$. It should be
noted that the agreement with  previous results should be taken qualitatively,
because the  metallicity cut in  different studies has been  chosen differently.
In addition, the target  selection in the ARGOS survey is  such that many more
metal  poor (halo)  stars are  included.  GIBS  data allows  us to  extend the
investigation  to  inner regions,  showing  that  the  trend of  sigma  versus
metallicity is inverted at $b=-1^\circ$, at  least along the bulge minor axis.
Figure~\ref{sig_bcost}  best  illustrates  this  point.   It  shows  that  for
$|b|>3^\circ$ the  metal poor  component has  a higher  sigma compared  to the
metal  rich one,  with  the  difference becoming  larger  moving outward.   At
latitude $b=-2^\circ$  the two  metallicity components have  comparable sigma,
while  at $b=-1^\circ$  it is  the metal  rich component  that shows  a higher
sigma.  Unfortunately we  only have one field at this  latitude, but there are
432 targets in  it,  and  therefore the  velocity  dispersion measurement  is
robust.  An indication of the inversion in the sigma versus metallicity trend,
at   very    low   latitudes,   was    already   found   in   the    data   by
\citet[][]{babusiaux+14},  \citep[see  also][]{babusiaux+16}.  With  its  very
homogeneous target  selection, the GIBS survey  allows us to confirm  that the
inversion is real.  This observational result seems to be due to the fact that
the  dispersion  of the  metal  poor  stars  increases  from the  outer  bulge
($b=8^\circ$)  towards the  plane, although  slower  than that  of metal  rich
stars,  but then  it  stays constant  in the  three  strips at  $b=-3.5^\circ,
-2^\circ, -1^\circ$. In other words, the velocity dispersion of the metal rich
component has  a gradient with latitude  much stronger than that  of the metal
poor component.

\begin{table}
\caption{Slope of V$_{\rm GC}$ vs longitude, in 5 latitude strips.}
\label{slopes}
\centering
{\tiny
\begin{tabular}{lrrrr}
\hline\hline
Latitude        &  [Fe/H]$<$$-0.3$ &  [Fe/H]$<$$-0.1$ &  [Fe/H]$>$$+0.1$ &  [Fe/H]$>$$+0.2$ \\
\hline
b$\sim$$-2^\circ$   &  8.2$\pm$1.4 &  9.3$\pm$1.1 & 10.9$\pm$1.0 & 10.8$\pm$1.1\\
b$\sim$$-3.5^\circ$ & 12.4$\pm$3.2 & 11.7$\pm$2.6 & 17.7$\pm$1.2 & 19.1$\pm$1.5\\
b$\sim$$+4.5^\circ$ &  8.6$\pm$1.3 &  9.6$\pm$1.1 &  8.3$\pm$1.0 &  9.2$\pm$1.1\\
b$\sim$$-6^\circ$   &  7.8$\pm$1.0 &  8.4$\pm$0.8 & 10.6$\pm$1.0 & 10.8$\pm$1.1\\
b$\sim$$-8.5^\circ$ &  7.0$\pm$0.7 &  7.4$\pm$0.6 &  8.7$\pm$0.8 &  8.3$\pm$0.8\\
\hline\hline
\end{tabular}}
\end{table}

\begin{figure}[h]
\centering
    {\includegraphics[angle=0,width=8.5cm]{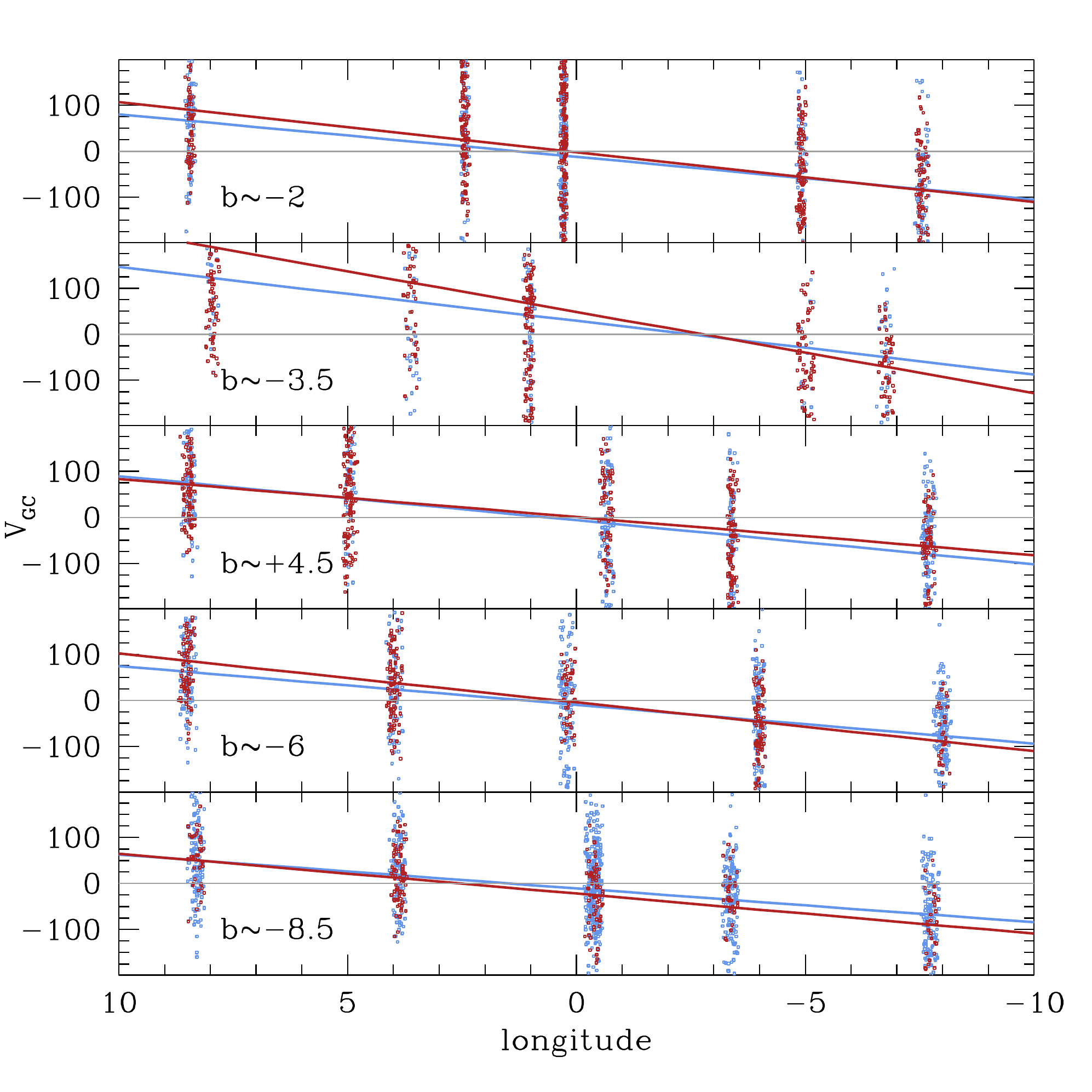}}
    \caption{Galactocentric  radial  velocity   versus  longitude,  at
      $\sim$ fixed latitudes,  as indicated in the  labels. Metal poor
      stars  ([Fe/H]$<-0.1$)  are  shown  in blue,  metal  rich  stars
      ([Fe/H]$>+0.1$) in  red. A  linear relation  has been  fit in
      each panel,  separately for  each metallicity  component.  Metal
      poor stars rotate marginally slower than metal rich ones.}
    \label{rv}
\end{figure}

\citet[][]{kunder+15} recently suggested that  bulge RR Lyrae, tracing
the oldest, more  metal-poor component, rotate slower than  RGB and RC
stars.   We investigate  whether this  signature is  imprinted in  the
present data as well.  Figure~\ref{rv} shows the galactocentric radial
velocity for all the stars, as a function of longitude, at fixed
  latitudes. Metal poor stars ([Fe/H]$<-0.1$) are shown in blue, metal
  rich stars ([Fe/H]$>+0.1$) in red.  A linear relation, including the
  errors in radial velocity, was fit to the data in each panel, for
  the two metallicity components  individually.  Metal poor stars show
  a marginally slower  rotation, i.e., a flatter  slope.  Adopting the
  metallicity cuts mentioned above, and  used elsewhere in this paper,
  the slopes  are always compatible  within the errors. However  if we
  move the  metal poor cut  at $-0.3$ dex, and  the metal rich  one at
  $+0.2$,   i.e.,   we   exclude   a  larger   metallicity   bin,   as
  cross-contaminated,  then  the  difference between  the  two  slopes
  become  more  significant  (Table~\ref{slopes}).   In  other  words,
  although the  signal is  marginally significant  in these  data, the
  GIBS  sample confirms  the findings  by \citet[][]{kunder+15},  that
  metal rich stars rotate faster than metal poor ones.

\begin{table*}
\caption{Components of the MDF fits and Nr of RC stars in each fields.}
\label{table_mdfs}
\centering
{\small
\begin{tabular}{l c c r c c r c c r}
\hline\hline
Field name & $<$[Fe/H]$>$ & $\sigma$ & N/N$_{\rm tot}$ & $<$[Fe/H]$>$ & $\sigma$ & N/N$_{\rm tot}$ & $<$[Fe/H]$>$ & $\sigma$ & N/N$_{\rm tot}$ \\
           & \multicolumn{3}{|c|}{Halo} & \multicolumn{3}{|c|}{Metal poor} & \multicolumn{3}{|c|}{Metal rich} \\ 
\hline
    LRp8p4 &  $-$1.38 &  0.08 &  0.02  & $-$0.30 &  0.31 &  0.54  &  0.24 &  0.17 &  0.44  \\
    LRp5p4 &   --     &   --  &  0.00  &  0.04   &  0.34 &  0.27  &  0.49 &  0.09 &  0.73  \\
    LRp0p4 &  $-$1.14 &  0.08 &  0.02  & $-$0.31 &  0.27 &  0.56  &  0.19 &  0.18 &  0.42  \\
    LRm3p4 &  $-$1.05 &  0.15 &  0.03  & $-$0.35 &  0.23 &  0.34  &  0.24 &  0.24 &  0.63  \\
    LRm8p4 &  $-$0.91 &  0.13 &  0.02  & $-$0.47 &  0.20 &  0.47  &  0.16 &  0.21 &  0.50  \\
&&&&&&&&& \\
    LRp0m1 &  $-$1.83 &  0.25 &  0.00  & $-$0.35 &  0.28 &  0.53  &  0.22 &  0.18 &  0.46  \\
&&&&&&&&& \\
    LRp8m2 &   --     &   --  &  0.00  & $-$0.87 &  0.12 &  0.03  &  0.00 &  0.34 &  0.97  \\
    LRp3m2 &  $-$1.40 &  0.16 &  0.01  & $-$0.23 &  0.24 &  0.43  &  0.36 &  0.18 &  0.56  \\
    LRp0m2 &  $-$0.95 &  0.13 &  0.04  & $-$0.26 &  0.29 &  0.56  &  0.37 &  0.19 &  0.40  \\
    LRm5m2 &   --     &   --  &  0.00  & $-$0.30 &  0.20 &  0.35  &  0.24 &  0.17 &  0.65  \\
    LRm8m2 &  $-$1.42 &  0.12 &  0.01  & $-$0.29 &  0.24 &  0.43  &  0.17 &  0.17 &  0.56  \\
&&&&&&&&& \\
    HRp8m3 &   --     &   --  &  0.00  & $-$0.45 &  0.21 &  0.17  &  0.24 &  0.19 &  0.83  \\
    HRp4m3 &  $-$0.89 &  0.26 &  0.05  & $-$0.18 &  0.19 &  0.41  &  0.32 &  0.15 &  0.54  \\
        BW &   --     &   --  &  0.00  & $-$0.23 &  0.39 &  0.48  &  0.33 &  0.16 &  0.52  \\
    HRm5m3 &  $-$1.60 &  0.06 &  0.01  & $-$0.39 &  0.25 &  0.16  &  0.21 &  0.16 &  0.83  \\
    HRm7m4 &  $-$1.53 &  0.05 &  0.01  & $-$0.45 &  0.16 &  0.29  &  0.22 &  0.21 &  0.70  \\
&&&&&&&&& \\
    LRp8m6 &  $-$1.27 &  0.30 &  0.03  & $-$0.27 &  0.21 &  0.52  &  0.29 &  0.20 &  0.45  \\
    LRp4m6 &  $-$0.82 &  0.09 &  0.03  & $-$0.26 &  0.22 &  0.40  &  0.28 &  0.22 &  0.57  \\
    bulge6 &  $-$0.90 &  0.31 &  0.09  & $-$0.35 &  0.17 &  0.41  &  0.08 &  0.16 &  0.50  \\
    LRm4m6 &  $-$0.77 &  0.15 &  0.06  & $-$0.35 &  0.17 &  0.41  &  0.23 &  0.22 &  0.52  \\
    LRm8m6 &   --     &   --  &  0.00  & $-$0.36 &  0.36 &  0.94  &  0.26 &  0.17 &  0.06  \\  
&&&&&&&&& \\
    LRp8m8 &  $-$1.00 &  0.06 &  0.02  & $-$0.34 &  0.24 &  0.80  &  0.32 &  0.09 &  0.18  \\
    LRp4m8 &  $-$1.25 &  0.09 &  0.01  & $-$0.35 &  0.26 &  0.59  &  0.31 &  0.17 &  0.41  \\
    LRp0m8 &  $-$0.85 &  0.54 &  0.04  & $-$0.37 &  0.22 &  0.81  &  0.33 &  0.17 &  0.16  \\
    LRm3m8 &  $-$1.01 &  0.13 &  0.05  & $-$0.35 &  0.24 &  0.82  &  0.36 &  0.11 &  0.13  \\
    LRm8m8 &  $-$1.01 &  0.18 &  0.07  & $-$0.41 &  0.18 &  0.65  &  0.13 &  0.27 &  0.28  \\
\hline
\multicolumn{10}{l}{Global MDFs} \\
 $b=-1^\circ$   & $-$1.83 & 0.25 & $<$0.01 & $-$0.35 & 0.28 & 0.53 &  0.22 & 0.18 & 0.46 \\
 $b=-2^\circ$   & $-$0.80 & 0.16 & $<$0.02 & $-$0.24 & 0.27 & 0.49 &  0.28 & 0.21 & 0.49 \\ 
 $b=-3.5^\circ$ & $-$1.44 & 0.15 & $<$0.01 & $-$0.31 & 0.31 & 0.33 &  0.26 & 0.20 & 0.67 \\ 
 $b=+4^\circ$   & $-$1.12 & 0.20 & $<$0.02 & $-$0.35 & 0.27 & 0.45 &  0.24 & 0.22 & 0.54 \\
 $b=-6^\circ$   & $-$0.87 & 0.36 & $<$0.06 & $-$0.32 & 0.25 & 0.51 &  0.22 & 0.23 & 0.43 \\
 $b=-8.5^\circ$ & $-$0.95 & 0.16 & $<$0.04 & $-$0.38 & 0.22 & 0.74 &  0.30 & 0.19 & 0.23 \\
\hline\hline
\end{tabular}}
\end{table*}

\section{Conclusions}

We presented  the metallicity of $\sim$5500  RC stars in 26  fields across the
Galactic bulge. In all but two fields, the metallicity distribution is clearly
bimodal with a  metal poor component centered  at [Fe/H]$_{\rm MP}\approx-0.4$
dex and a  metal rich component centered at  [Fe/H]$_{\rm MR}\approx+0.3$ dex.
A small (few  $\%$) population of stars with [Fe/H]$<-0.8$  is also identified
in most  of the  fields, tentatively interpreted  as halo  contamination.  The
latter population is highly incomplete,  because our target selection box does
not extend to the blue of the RC, therefore we did not attempt to characterize
it.

\begin{figure}[h]
\centering
    {\includegraphics[angle=0,width=8cm]{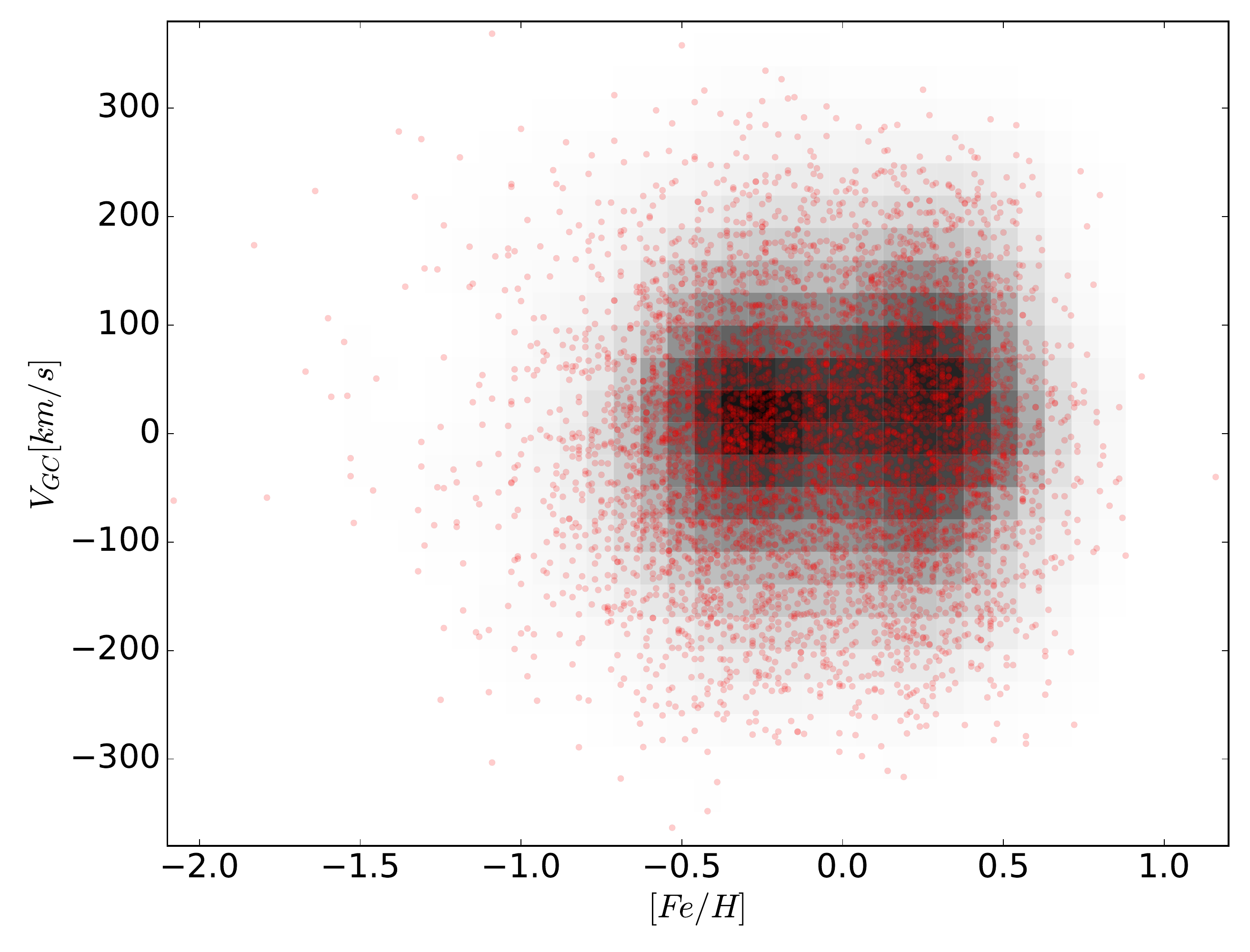}}
    \caption{Metallicity versus galactocentric radial velocity for the whole
sample discussed here. The two metallicity components, that we call metal 
rich and metal poor, respectively, are clearly seen here.}
    \label{rv_fe_tot}
\end{figure}

Figure~\ref{rv_fe_tot}  shows  the  metallicity versus  galactocentric  radial
velocity  for  all  the  stars  in the  present  sample.  Two  populations  in
metallicity  are clearly  seen  here.   No stars  were  measured outside  this
figure's box, i.e., we do not  detect high velocity stars, inconsistent with
the tail  of the main  velocity distribution.   The two main  bulge components
peak rougly  at the  same metallicity  in different  fields, but  the relative
fraction of stars in each of  them changes dramatically across the bulge area,
giving rise to  well known radial gradients in the  bulge mean metallicity. In
the outer region studied here ($b=-8.5^\circ$), the metal poor component largely
outnumbers the  metal rich one,  with the former  accounting to 73$\%$  of the
total  number of  targets observed  at this  latitude, and  the latter  to the
remaining 27$\%$. The relative fraction of metal poor stars drops to 33$\%$ of
the total  at $b=-3.5^\circ$: a  result in  very good agreement  with previous
findings   by    the   ARGOS    \citep[][]{ness+13mdf}   and    the   Gaia-ESO
\citep[][]{rojas-arriagada+14} surveys.

The present dataset allows us to  reach latitudes closer to the Galactic plane
than previous  studies, probing the population  of the inner bulge.   A first,
important result  of the present work  is that the relative  fraction of metal
poor   stars  increases   again  from   $b=-3.5^\circ$  to   $b=-2^\circ$  and
$b=-1^\circ$,  reaching 49$\%$  and  53$\%$ of  the total  in  the latter  two
fields, respectively.

Having demonstrated that the bulge  includes two metallicity components whose
relative fraction changes across the bulge  area, i.e., they have a different
spatial distribution,  we mapped their  density distribution by  coupling the
relative fractions estimated here with  the bulge stellar density map derived
in   \citet[][]{valenti+16}.   The   resulting   density   maps,   shown   in
Fig.~\ref{metal-density},  demonstrate that  the metal  poor component  has a
spheroid-like spatial distribution, versus a boxy distribution of the metal rich
component,  and  the  radial  density  gradient of  the  metal  poor  one  is
steeper. Due  to the coarse spatial  sampling of the GIBS  fields, these maps
have been  derived imposing  4-fold simmetry (about  the Galactic  plane, and
about  the bulge  projected minor  axis). Therefore,  we are  by construction
unable to identify possible asymmetries related  to the fact that at positive
(/negative) longitudes we see the near (/far) side of the bar.

The two bulge metallicity components  not only have different mean metallicity
and different  spatial distribution, they  also have different  kinematics. As
already  found  by  the  BRAVA   \cite[][]{kunder+12}  and  the  ARGOS  survey
\citep[][]{ness+13kin},  in the  outer  bulge ($|b|>4^\circ$)  the metal  poor
component has a  higher radial velocity dispersion compared to  the metal rich
one, at  all longitudes.  A  new result of  the present investigation  is that
this   behaviour  is   reversed  in   the  inner   bulge.   Specifically,   at
$b=-3.5^\circ$  and $b=-2^\circ$  the velocity  dispersion of  the metal  poor
stars becomes very similar to that of the metal rich ones, and at $b=-1^\circ$
it becomes smaller. This is due to  the fact that the velocity distribution of
the  metal poor  component has  a mild  gradient with  latitude, ranging  from
$\sigma\sim$80   km/s   at   $b=-8.5^\circ$   to   $\sigma\sim$125   km/s   at
$b=-1^\circ$. On  the contrary, the  metal rich  component has a  low velocity
dispersion in  the outer  bulge ($\sigma\sim$60  km/s at  $b=-8.5^\circ$) that
grows up to $\sigma\sim$145 km/s at  $b=-1^\circ$. In other words, the central
{\it sigma-peak} discovered in Paper~I, is mostly due to the metal rich stars.
The origin of this {\it sigma-peak}, in the inner $\sim$250 pc of the Galactic
bulge, has  been investigated by \citet[][]{valenti+16}  who demonstrated that
it coincides with a peak in the stellar density (traced by counting the number
of RC  stars across  the VVV survey  area) and  thus with a  peak in  the mass
density.

Finally, we  investigated whether  the GIBS  data show  any evidence  for the
metal poor component rotating slower than the metal rich one, as suggested by
\citet[][]{kunder+15},  who traced  the former  by means  of RR  Lyrae stars.
Indeed, the metal poor component shows  a marginally slower rotation also in
the present dataset, although the difference is not compelling here.

Summarising, we unambiguously detect two components in the Galactic bulge,
each having a relatively narrow metallicity distribution, so that the global
bulge MDF is clearly bimodal. These two components have a different spatial
distribution, with the metal poor being more centrally concentrated, and with
a rounder density distribution as projected in the sky.  On the contrary, the
metal rich component is clearly boxy, as expected for a bar seen edge on. We
stress that, although the present data are consistent with a spheroidal
distribution of the metal poor component (although we did not investigate the
3rd component, along the line of sight), we are not presenting evidence in
favor of it having a different origin (i.e., {\it a classical bulge}).
Different models of bulge formation are being developed, arguing that a
spheroidal shape can be obtained through different formation scenarios
(Debattista, {\it priv.  comm.}).  We provide here new observational
constraints that were not available so far, and need to be fullfilled by any
Milky Way bulge formation model.  
   

\begin{acknowledgements}
We gratefully  acknowledge support by  the Ministry of  Economy, Development,
and Tourism's  Millennium Science Initiative through  grant IC120009, awarded
to  The  Millennium Institute  of  Astrophysics  (MAS), by  Fondecyt  Regular
1150345  and  by  the  BASAL-CATA  Center  for  Astrophysics  and  Associated
Technologies  PFB-06.   This work  was  finalized  at  the Aspen  Center  for
Physics,   which  is   supported   by  National   Science  Foundation   grant
PHY-1066293. MZ  and DM were partially  supported by a grant  from the Simons
Foundation, during their stay in Aspen.
\end{acknowledgements}

\bibliographystyle{aa}


\end{document}